\documentclass[12pt,preprint]{aastex}
\newcommand{\kms}{km s$^{-1}$}
\newcommand{\zabs}{$z_{\rm abs}$}
\newcommand{\lya}{Ly$\alpha$\ }

\newcommand{\nav}{$N_{\rm a}(v)$}
\newcommand{\vlsr}{$v_{\rm LSR}$}
\slugcomment{Submitted to the {\bf Astronomical Journal}}
\shortauthors{TRIPP et al.}
\shorttitle{Nature of Complex C}
\begin{document}

\title{Complex C: A Low-Metallicity High-Velocity Cloud 
Plunging into the Milky Way\altaffilmark{1}}

\altaffiltext{1}{Based on observations with the NASA/ESA 
{\it Hubble Space Telescope}, obtained at the Space 
Telescope Science Institute, which is operated by the 
Association of Universities for Research in Astronomy, 
Inc., under NASA contract NAS 5-26555.}

\author{Todd M. Tripp,\altaffilmark{2,3} Bart P. 
Wakker,\altaffilmark{4} Edward B. Jenkins,\altaffilmark{2} 
C. W. Bowers,\altaffilmark{5} A. C. Danks ,\altaffilmark{5} 
R. F. Green,\altaffilmark{6} S. R. Heap,\altaffilmark{5} C. 
L. Joseph,\altaffilmark{7} M. E. Kaiser,\altaffilmark{8} J. 
L. Linsky,\altaffilmark{9} and B. E. 
Woodgate\altaffilmark{5}}

\altaffiltext{2}{Princeton University Observatory, 
Peyton Hall, Princeton, NJ 08544, 
Electronic mail: tripp@astro.princeton.edu}

\altaffiltext{3}{Department of Astronomy, University of Massachusetts, 
Amherst, MA 01003}

\altaffiltext{4}{Department of Astronomy, University of Wisconsin, 475 
N. Charter St., Madison, WI 53706}

\altaffiltext{5}{NASA Goddard Space Flight Center, Code 681, Greenbelt, 
MD 20771}

\altaffiltext{6}{ National Optical Astronomy Observatories, Tucson, AZ 
85726}

\altaffiltext{7}{Department of Physics and Astronomy, Rutgers 
University, New Brunswick, NJ 08855}

\altaffiltext{8}{Department of Physics and Astronomy, Johns Hopkins 
University, 3400 North Charles Street, Baltimore, MD 21218}

\altaffiltext{9}{JILA, University of Colorado and NIST, Boulder, CO 
80309}

\begin{abstract}
\small
  We present evidence that high-velocity cloud (HVC) Complex C is a
  low-metallicity gas cloud that is plunging toward the disk and
  beginning to interact with the ambient gas that surrounds the Milky
  Way.  This evidence begins with a new high-resolution (7 \kms\ FWHM)
  echelle spectrum of 3C 351 obtained with the Space Telescope Imaging
  Spectrograph (STIS). 3C 351 lies behind the low-latitude edge of
  Complex C, and the new spectrum provides accurate measurements of
  \ion{O}{1}, \ion{Si}{2}, \ion{Al}{2}, \ion{Fe}{2}, and \ion{Si}{3}
  absorption lines at the velocity of Complex C; \ion{N}{1},
  \ion{S}{2}, \ion{Si}{4}, and \ion{C}{4} are not detected at
  $3\sigma$ significance in Complex C proper.  However, \ion{Si}{4}
  and \ion{C}{4} as well as \ion{O}{1}, \ion{Al}{2}, \ion{Si}{2} and
  \ion{Si}{3} absorption lines are clearly present at somewhat higher
  velocities associated with a ``high-velocity ridge'' (HVR) of 21cm
  emission. This high-velocity ridge has a similar morphology to, and
  is roughly centered on, Complex C proper. The similarity of the
  absorption lines ratios in the HVR and Complex C suggest that these
  structures are intimately related. In Complex C proper we find [O/H]
  = $-0.76^{+0.23}_{-0.21}$.  For other species, the measured column
  densities indicate that ionization corrections are important. We use
  collisional and photoionization models to derive ionization
  corrections; in both models we find that the overall metallicity $Z
  = 0.1 - 0.3 Z_{\odot}$ in Complex C proper, but nitrogen must be
  underabundant. The iron abundance indicates that the Complex C
  contains very little dust. The size and density implied by the
  ionization models indicate that the absorbing gas is not
  gravitationally confined. The gas could be pressure-confined by an
  external medium, but alternatively we may be viewing the leading
  edge of the HVC, which is ablating and dissipating as it plunges
  into the Milky Way.  \ion{O}{6} column densities observed with {\it
    FUSE} toward nine QSOs/AGNs behind Complex C support this
  conclusion: $N$(\ion{O}{6}) is highest near 3C 351, and the
  \ion{O}{6}/\ion{H}{1} ratio increases substantially with decreasing
  latitude, suggesting that the lower-latitude portion of the cloud is
  interacting more vigorously with the Galaxy. The other sight lines
  through Complex C show some dispersion in metallicity, but with the
  current uncertainties, the measurements are consistent with a
  constant metallicity throughout the HVC. However, all of the Complex
  C sight lines require significant nitrogen underabundances. Finally,
  we compare the 3C 351 data to high-resolution STIS observations of
  the nearby QSO H1821+643 to search for evidence of outflowing
  Galactic fountain gas that could be mixing with Complex C. We find
  that the intermediate-velocity gas detected toward 3C 351 and
  H1821+643 has a higher metallicity and may well be a
  fountain/chimney outflow from the Perseus spiral arm.  However, the
  results for the higher-velocity gas are inconclusive: the HVC
  detected toward H1821+643 near the velocity of Complex C could have
  a similar metallicity to the 3C 351 gas, or it could have a
  significantly higher $Z$, depending on the poorly constrained
  ionization correction.
\end{abstract}


\section{Introduction}

The Galactic high-velocity clouds (HVCs), gas clouds detected via 21 cm 
emission or UV/optical absorption at velocities that deviate 
substantially from normal Galactic rotation (Wakker \& van Woerden 
1997), have potentially important implications regarding the structure 
and evolution of the Galaxy and Local Group. For example, if the HVCs 
provide a sufficient quantity of infalling low-metallicity gas, they 
can alleviate the 
G-dwarf problem, the long-standing discrepancy between the observed 
metallicity distribution of G-dwarf stars and theoretical expectations 
(e.g., Larson 1972; Tosi 1988; Wakker et al. 1999; Gibson et al. 2002). 
The HVCs may have important cosmological implications as well. 
Hierarchical models of galaxy formation within the cold dark matter 
framework predict many more dwarf satellite galaxies within the Local 
Group than have been detected/identified (e.g., Klypin et al. 1999; 
Moore et al. 1999). A variety of solutions to this problem have been 
proposed, including the possibility that the HVCs are the missing 
dwarfs which, for some reason, have not formed readily detectable stars 
(e.g., Blitz et al. 1999; Klypin et al. 1999; Gibson et al. 2002).

However, the nature of most high-velocity gas is still poorly 
understood, mainly because the cloud distances are highly uncertain. 
Some of the HVCs are clearly material stripped out of the Magellanic 
Clouds, the ``Magellanic Stream'' (Mathewson, Cleary, \& Murray 1974; 
Putman et al. 1998; Lockman et al. 2002), and some high-velocity 
absorption lines observed towards disk stars are most likely related to 
the interaction between stars/supernovae and the ISM (e.g., Cowie, 
Songaila, \& York 1979; Trapero et al. 1996; Jenkins et al. 1998, 2000; 
Welty et al. 1999; Tripp et al. 2002). Apart from these clouds, direct 
distance constraints are scarce and difficult to obtain (Wakker 2001), 
and consequently a variety of HVC models remain viable. The HVCs may 
have a galactic origin such as the Galactic fountain (Shapiro \& Field 
1976; Bregman 1980), or they may be extragalactic (e.g., Oort 1970; 
Blitz et al. 1999; Braun \& Burton 1999).

Given the difficulty of direct distance measurements, it is important
to explore other constraints on the nature of these objects. For
example, constraints on the distribution and size of the HVCs can be
derived from studies of other galaxy groups using either QSO
absorption line statistics (Charlton, Churchill, \& Rigby 2000) or
surveys for redshifted 21 cm emission (Zwaan \& Briggs 2000; Zwaan
2001; Braun \& Burton 2001; Pisano \& Wilcots 2003). For the Milky Way
HVCs, it has been suggested that the H$\alpha$ emission from the
clouds constrains their distances (e.g., Bland-Hawthorn et al. 1998).
If the gas is photoionized by UV flux escaping from the Galaxy, then
nearby clouds should be substantially brighter in H$\alpha$ than
distant HVCs. H$\alpha$ emission has in fact been detected from a
variety of HVCs (e.g., Weiner \& Williams 1996; Tufte et al. 1998,
2002; Weiner, Vogel, \& Williams 2002), and detailed photoionization
models place the clouds roughly 5 $-$ 50 kpc away based on the
observed H$\alpha$ intensities (Weiner et al. 2002; Bland-Hawthorn \&
Maloney 2002). However, several observations cloud the interpretation
of H$\alpha$ intensities. First, the Magellanic Stream, which has an
independently constrained distance, is much brighter in H$\alpha$ than
predicted by the photoionization models (Weiner et al. 2002;
Bland-Hawthorn \& Maloney 2002). Second, the Magellanic Stream
H$\alpha$ emission is spatially variable (see Figure 1 in Weiner et
al. 2002), also contrary to the predictions of the photoionization
models. Finally, \ion{O}{6} absorption has been detected in a
substantial fraction of the HVCs (Sembach et al. 2000, 2002; Wakker et
al. 2002). While \ion{O}{6} can be produced by photoionization in
large, low-density intergalactic gas clouds (e.g., Tripp \& Savage
2000; Tripp et al. 2001), the \ion{O}{6} in HVCs is almost certainly
collisionally ionized (the cloud sizes required by photoionization
models are excessive for HVCs, see Sembach et al.  2002). These
observations suggest that collisional processes play an important role
in the ionization of HVCs and the production of H$\alpha$ emission.
The interaction of the rapidly moving clouds with ambient magnetic
fields may create instabilities that also play a role in the gas
ionization and production of H$\alpha$ emission (Konz et al.  2001).
For these reasons, H$\alpha$ distance constraints from photoionization
models may need to be revisited.

Ultraviolet absorption lines in the spectra of background quasars and 
active galactic nuclei (AGNs) provide another sensitive probe for the 
study of Galactic (as well as extragalactic) HVCs. UV absorption lines 
provide detailed information on abundances and physical conditions in 
the gas, and this in turn can be compared to predictions of various 
models. For example, it is expected that Galactic fountain gas would 
have a substantially higher metallicity than infalling extragalactic 
gas or gas stripped from a satellite galaxy. {\it Relative} metal 
abundance patterns may also provide insights on the enrichment history 
of the HVCs; if the gas is relatively pristine, overabundances of 
$\alpha$ elements or underabundances of nitrogen might be observed, as 
seen in low-metallicity stars (McWilliam 1997) and \ion{H}{2} regions 
(Vila-Costas \& Edmunds 1993; Henry, Edmunds, \& K\"{o}ppen 2000). 

In this paper we use the absorption line technique to explore the 
nature of HVC Complex C, a large HVC (roughly 20$^{\circ} \times 
90^{\circ}$) that is at least $\sim$5 kpc away (van Woerden et al. 
1999). A number of Complex C abundance measurements\footnote{Throughout 
this paper we express abundances with the usual logarithmic notation, 
[X/Y] = log(X/Y) $-$ log(X/Y)$_{\odot}$, and we indicate the overall 
metallicity with the variable $Z$.} have been published for a variety 
of species ranging from [N/H] = $-1.94$ to [Fe/H] = $-0.3$ (Bowen, 
Blades, \& Pettini 1995; Wakker et al. 1999; Murphy et al. 2000; 
Richter et al. 2001; Gibson et al. 2001; Collins et al. 2002). However, 
some of these abundances may be confused by ionization effects. Species 
such as \ion{S}{2} or \ion{Fe}{2} can arise in ionized as well as 
neutral gas leading to {\it overestimates} of [S/H] or [Fe/H] if no 
ionization correction is applied, while other species such as 
\ion{N}{1} can be more readily ionized than \ion{H}{1} leading to {\it 
underestimation} of the elemental abundance. The most robust species 
for constraining the metallicity of HVCs is \ion{O}{1}. Like sulfur, 
oxygen is only lightly depleted by dust grains (see \S 6.2.1 in Moos et 
al. 2002, and references therein), but more importantly, the ionization 
potential of \ion{O}{1} is nearly identical to that of \ion{H}{1}, and 
\ion{O}{1} is strongly locked to \ion{H}{1} by resonant charge exchange 
(Field \& Steigman 1971). Consequently, oxygen abundances based on 
\ion{O}{1} and \ion{H}{1} are relatively impervious to ionization 
effects (unless the gas is quite substantially ionized). 

Richter et al. (2001) have presented the first measurement of [O/H] in 
Complex C using spectra of PG1259+593 obtained with the Space Telescope 
Imaging Spectrograph (STIS) and the {\it Far Ultraviolet Spectroscopic 
Explorer (FUSE)}; they find [O/H] = $-1.03^{+0.37}_{-0.31}$. Notably, 
Richter et al. also report that nitrogen is highly underabundant (by 
$\sim$ 1 dex compared to oxygen), and the $\alpha$ elements are 
marginally overabundant compared to iron. These results suggest that 
Complex C is a relatively pristine extragalactic cloud plunging into 
the Galaxy for the first time. This notion has been challenged by 
Gibson et al. (2001), who have observed \ion{S}{2} in several 
directions through Complex C and find evidence of spatial variability 
of the sulfur abundance. Sulfur abundances derived from \ion{S}{2} 
alone are prone to ionization effects, as noted, but recently Collins, 
Shull, \& Giroux (2003) have also reported variable metallicity in 
Complex C based on oxygen measurements.

Here we present new constraints on the metallicity as well 
as the physical conditions, structure, and nature of 
high-velocity cloud Complex C from UV absorption lines. Our analysis is 
mainly based on high-resolution echelle spectroscopy of 3C 351 obtained 
with STIS, but we make use of high-resolution observations of nearby 
sight lines to augment the results with spatial information, e.g., on 
the transverse extent of the absorbing gas. We also briefly discuss 
intermediate-velocity gas in the 3C 351 direction. The paper is 
organized as follows: We begin with a summary of 21cm emission 
observations of high-velocity gas in the direction of 3C 351 (\S 
\ref{sec21cm}), followed by a brief summary of the STIS observations 
(\S \ref{secobs}) and the absorption line measurements (\S 
\ref{secmeas}). We then argue that ionization corrections are likely to 
be important for these clouds, and we employ collisional and 
photoionization models to constrain their abundances (\S 
\ref{secabun}). The abundances indicate that Complex C and the 
intermediate-velocity gas likely have different origins; we suggest 
that Complex C is an infalling, low-metallicity cloud while the IVC is 
outflowing gas from the Perseus spiral arm. In the discussion of our 
measurements, we consider the implications of the density and size of 
the absorbing gas derived from the ionization models including the 
confinement of the gas (\S \ref{secstructure}), and we compare 3C 351 
to other nearby sight lines in the vicinity of Complex C (\S 
\ref{seccompare}). We find evidence that the lower latitude portion of 
Complex C is more affected by interactions with the ambient medium that 
the higher latitude region of the cloud. We summarize our conclusions 
in \S \ref{secsum}. Throughout this paper we present spectra and 
velocities relative to the Local Standard of Rest (LSR).\footnote{We 
adopt the ``standard'' definition of the Local Standard of Rest 
(Delhaye 1965; Kerr \& Lynden-Bell 1986), in which the Sun is moving in 
the direction $l = 56^{\circ}, b = 23^{\circ}$ at 19.5 km s$^{-1}$. 
With this convention, the conversion to heliocentric velocity is given 
by $v_{\rm helio} = v_{\rm LSR} - 16.5$ km s$^{-1}$ for 3C 351.}

\section{High-Velocity Clouds Toward 3C 351\label{sec21cm}}

\begin{figure}
\epsscale{0.65}
\plotone{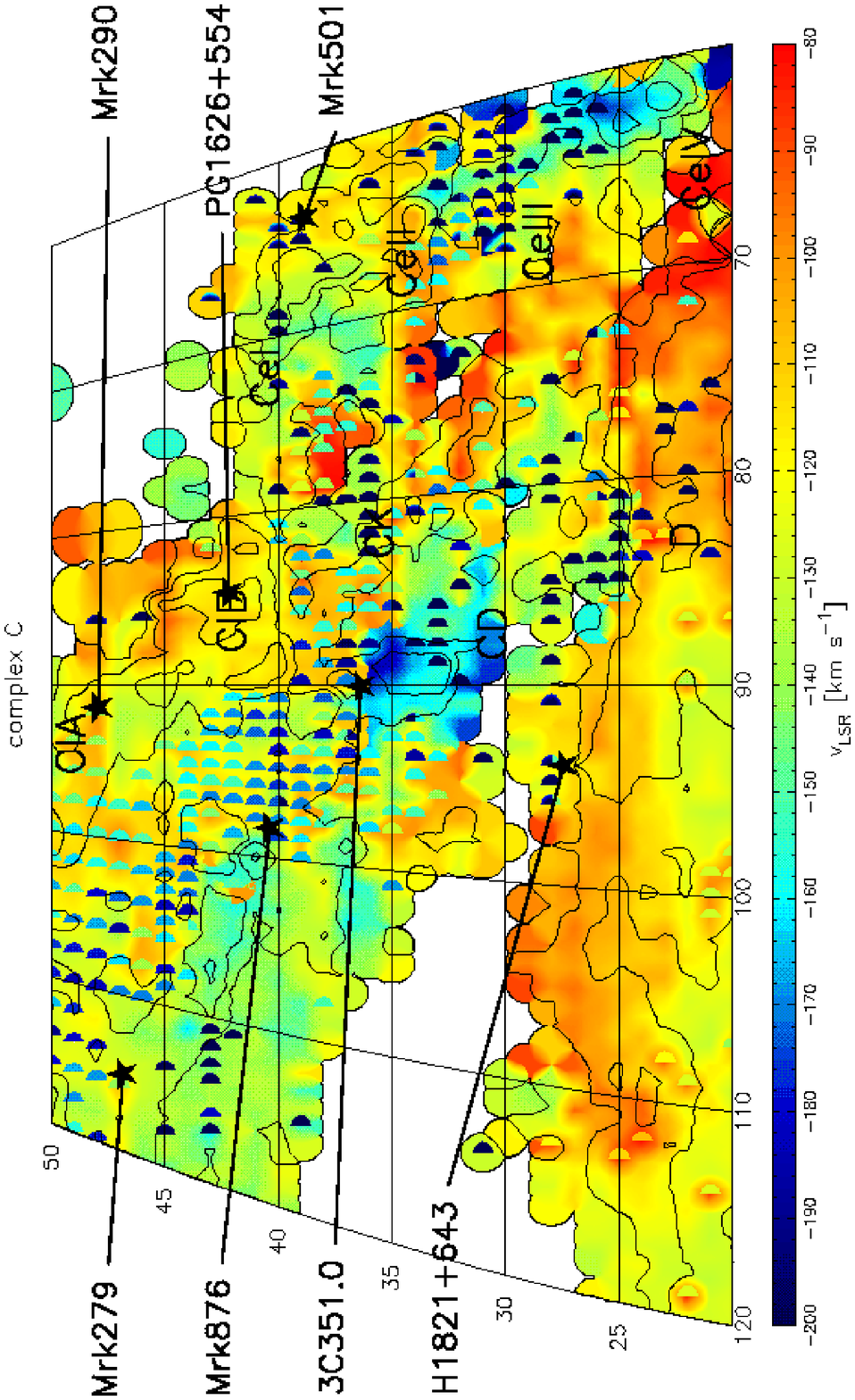}
\caption[]{\scriptsize Map of HVC Complex C 21 cm emission,
from Hulsbosch \& Wakker (1988), in Galactic coordinates
(longitude increases from right to left along the x-axis).
This map is centered on the sight line to 3C 351 and shows
only a portion of the cloud (for a map of the entire HVC,
see Figure 6 in Wakker 2001). Emission from high-velocity
gas in the Outer Arm is also evident in this map closer to
the plane; at $l > 90^{\circ}$ and $b < 30^{\circ}$, the
high-velocity emission is dominated by the Outer Arm. At
lower longitudes the distinction is less clear; the line
from $(l,b) = (70,25)$ to (88,29) roughly delineates these
structures. LSR velocities are indicated by color using the
scale at the bottom of the figure, and the contours show
brightness temperatures of 0.05, 0.4, and 1.0 K. The
half-circles show the high-velocity ridge at $v_{\rm LSR}
\approx -200$ \kms ; this higher velocity feature is
clearly detected in absorption toward 3C351 (see text and
Figures~\ref{oispec}-\ref{navprofs}). The emission cores
CIA,CIB, CeI-CeIV, CK, CD, and D are also labeled along
with other extragalactic sight lines of interest discussed
in this paper. \label{hw_map21cm}}
\end{figure}

\begin{figure}
\epsscale{0.65}
\plotone{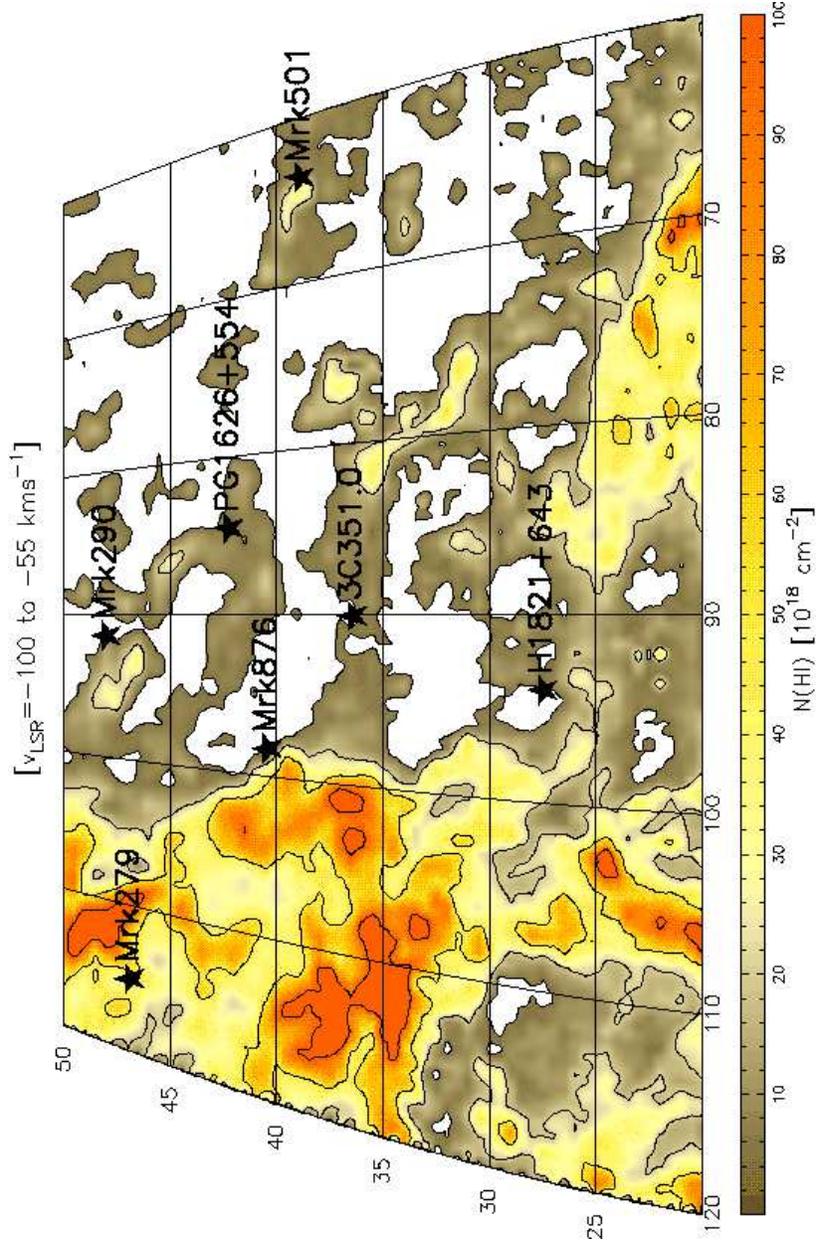}
\caption[]{Map of 21 cm emission from intermediate-velocity
gas in the vicinity of 3C 351 from the Leiden-Dwingeloo
survey including the following velocity range:
$-100 \leq v_{\rm LSR} \leq -55$ \kms . Emission in this
velocity range is dominated by the ``IV Arch'' at $l
\gtrsim 98^{\circ}$; emission at lower latitudes may be
associated with Complex K (see \S 4.24 and \S 4.27,
respectively, in Wakker 2001) or the Perseus Arm (see text,
\S~\ref{secouter}).\label{lds_ivc}}
\end{figure}

\begin{figure}
\epsscale{0.65}
\plotone{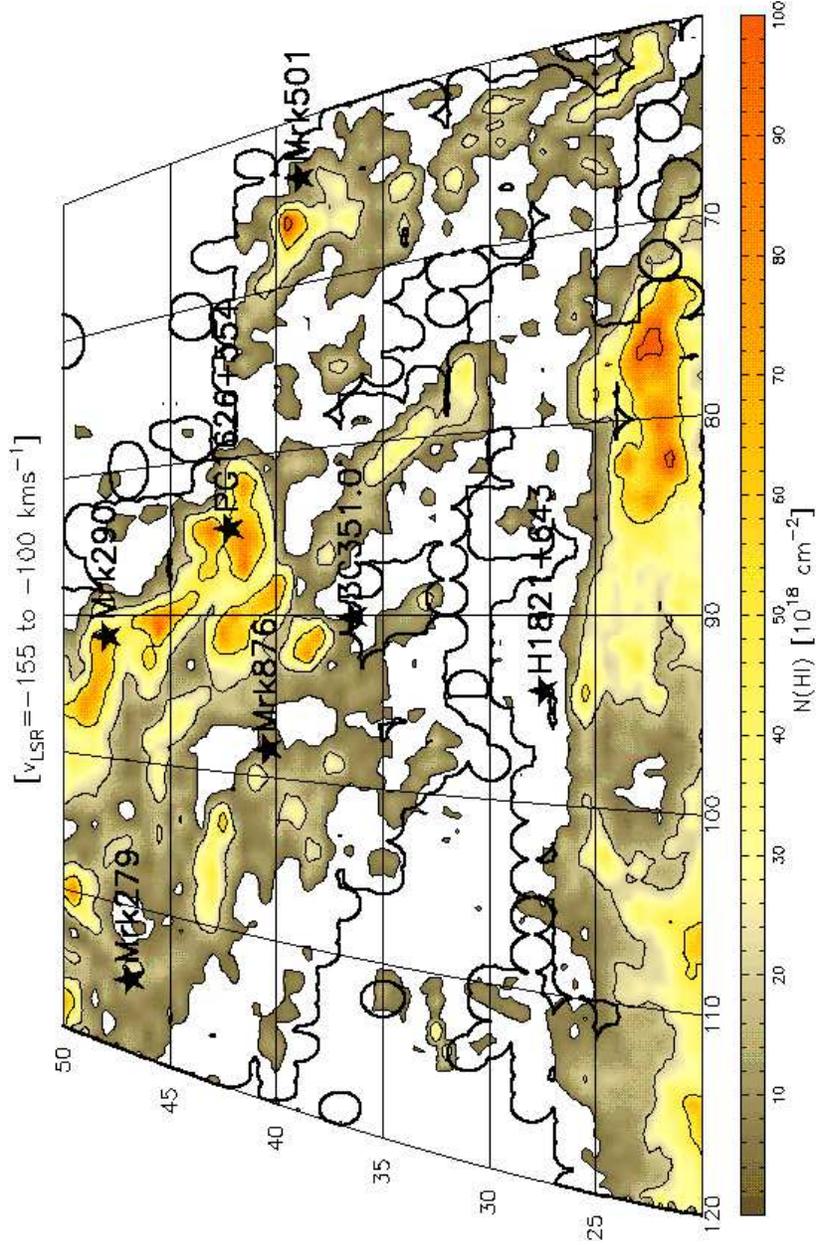}
\caption[]{Partial map of Complex C 21 cm emission as in
Figure~\ref{hw_map21cm}, but based on data from the
Leiden-Dwingeloo Survey (Hartmann \& Burton 1997). Emission
at $b \lesssim 27^{\circ}$ is mainly from the Outer Arm;
higher-latitude emission is associated with Complex C. In
this figure, colors indicate the \ion{H}{1} column density
as shown by the scale at the bottom, integrated from
$v_{\rm LSR} = -155$ to $-100$ \kms . The lowest contour
from the LDS data corresponds to the $3\sigma$ limit of
$\sim 5 \times 10^{18}$ cm$^{-2}$. The thick line indicates
the $N$(\ion{H}{1}) = $2 \times 10^{18}$ cm$^{-2}$ contour
from the Hulsbosch \& Wakker map shown in
Figure~\ref{hw_map21cm}.\label{lds_compc}}
\end{figure}

\begin{figure}
\epsscale{0.65}
\plotone{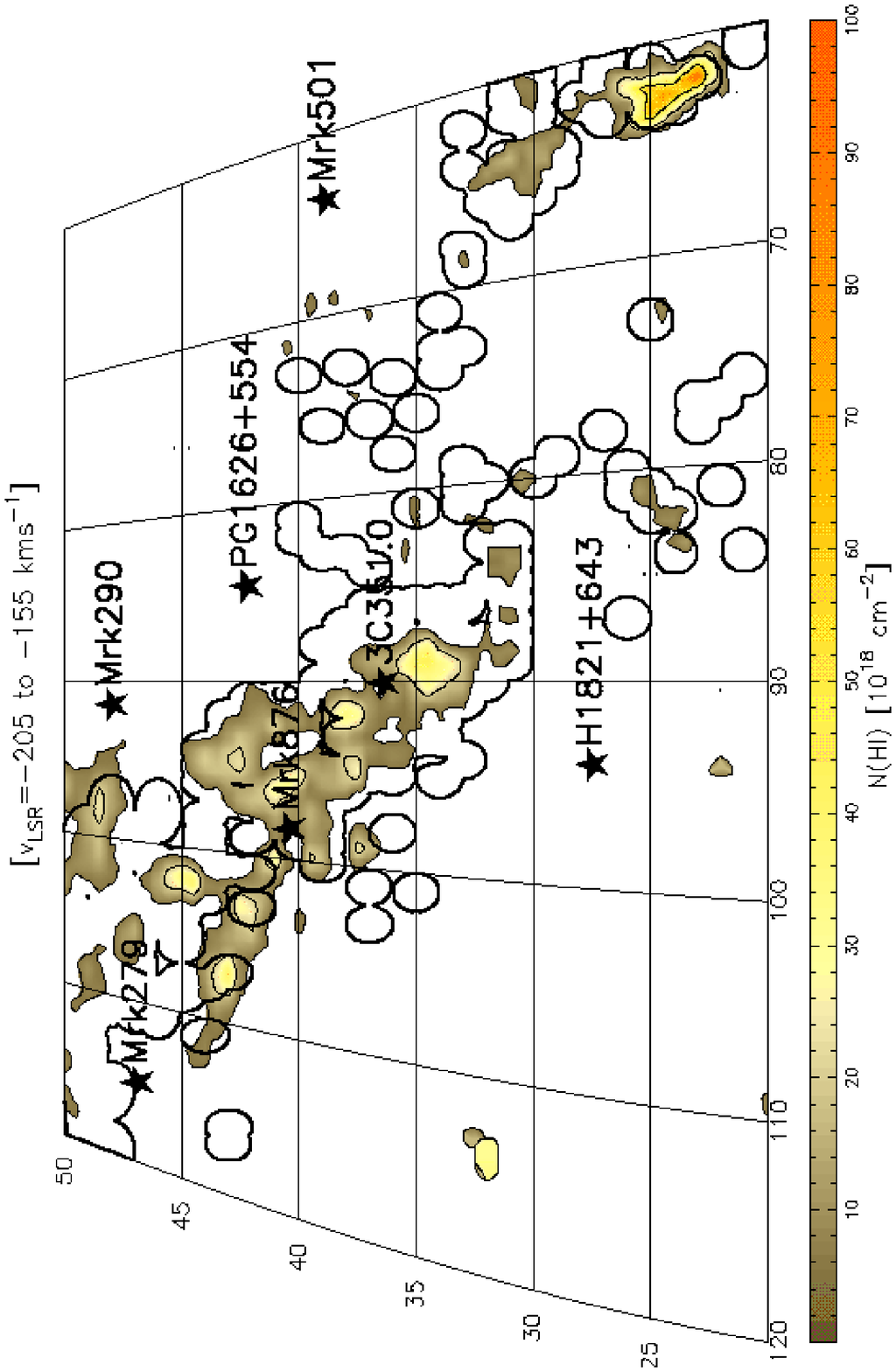}
\caption[]{Map of 21 cm emission from the ``High-Velocity
Ridge'' (see text) from the Leiden-Dwingeloo survey. This
figure is identical to Figure~\ref{lds_compc}, except the
emission is integrated over a different velocity range:
$-205 \leq v_{\rm LSR} \leq -155$ \kms .\label{lds_hvr}}
\end{figure}

The sight line to 3C 351 ($l = 90.08^{\circ}$, $b = +36.38^{\circ}$) 
probes an intriguing region of the 
high-velocity sky. A portion of the Hulsbosch \& Wakker (1988) 21 cm 
emission map of Complex C, centered on 3C 351, is shown in 
Figure~\ref{hw_map21cm}. In this figure, the color scale shows the LSR 
gas velocity while the contours indicate brightness temperature. The 
lowest contour shown corresponds to $N$(\ion{H}{1}) = $2 \times 
10^{18}$ cm$^{-2}$. Complex C shows several well-defined cores; the 3C 
351 sight line is roughly equidistant in projection from the CIB, CD, 
and CK cores [see Wakker (2001, and references therein) for definitions 
of the cores and nomenclature]. CK has a lower velocity than most of 
Complex C and may be more closely associated with the 
intermediate-velocity cloud Complex K than the high-velocity Complex C 
cloud (see Wakker 2001 and Haffner, Reynolds, \& Tufte 2001). 
Figure~\ref{lds_ivc} shows the intermediate-velocity 21 cm sky in the 
vicinity of 3C 351. The map in Figure~\ref{lds_ivc} plots \ion{H}{1} 
column densities derived from the 21 cm Leiden-Dwingeloo Survey (LDS; 
Hartmann \& Burton 1997) integrated over $-100 \leq v_{\rm LSR} \leq -
55$ \kms . At $l \gtrsim 100^{\circ}$, the intermediate-velocity sky is 
dominated by the ``IV Arch'', but at lower latitudes it is unclear 
whether the emission is associated with the IV Arch, Complex K, or a 
lower-velocity region of Complex C (see Wakker 2001). We provide 
evidence below that the CK IVC and Complex C proper have different 
origins (\S \S 5.4,6). We shall refer to the 3C 351 
intermediate-velocity absorption system as C/K hereafter and Complex C 
proper ($v_{\rm LSR} \approx -130$ km s$^{-1}$) as simply Complex C.

A notable feature of Complex C is the presence of 21 cm emission at 
substantially higher velocities ($v_{\rm LSR} \leq -170$ \kms ) along 
the central ridge of the cloud superimposed on the main HVC emission 
($-155 \lesssim v_{\rm LSR} \lesssim -80$ \kms ). The higher velocity 
emission is indicated with half-circles in Figure~\ref{hw_map21cm} (see 
also Figure 6 in Wakker 2001). The nature of the higher velocity 
component is not entirely clear, but it has a similar morphology to the 
lower velocity Complex C gas, and it is centered on Complex C. For 
convenience, we refer to this cloud as the ``high-velocity ridge'' 
(HVR) in this paper. A variety of UV absorption lines are detected from 
the high-velocity ridge in the 3C 351 spectrum as well as Complex C, 
and unlike C vs. C/K, the UV absorption lines provide evidence that the 
HV ridge and Complex C are closely related (see below).

The \ion{H}{1} column densities in Complex C and the high-velocity 
ridge are shown in Figures~\ref{lds_compc} and \ref{lds_hvr}, 
respectively. Figures~\ref{lds_compc}-\ref{lds_hvr} show 21 cm emission 
maps from the LDS integrated over $-155 \leq v_{\rm LSR} \leq -100$ 
\kms\  (Complex C proper) and $-205 \leq v_{\rm LSR} \leq -170$ \kms\ 
(high-velocity ridge). In these figures, the color scale and thin 
contours reflect $N$(\ion{H}{1}) according to the scale at the bottom. 
The last thin contour is drawn at the 3$\sigma$ limit of the LDS, 
$N$(\ion{H}{1}) = $5 \times 10^{18}$ cm$^{-2}$. The thick line shows 
the $2 \times 10^{18}$ cm$^{-2}$ contour from the more sensitive 
Hulsbosch \& Wakker map in Figure~\ref{hw_map21cm} integrated over the 
same velocity range. Much of the high-velocity emission detected closer 
to the plane 
in Figures~\ref{hw_map21cm}-\ref{lds_compc} is mainly associated with 
the Outer Arm and the warp of the outer Milky Way (Wakker 2001, and 
references therein). 

\begin{figure}
\epsscale{1.0}
\plotone{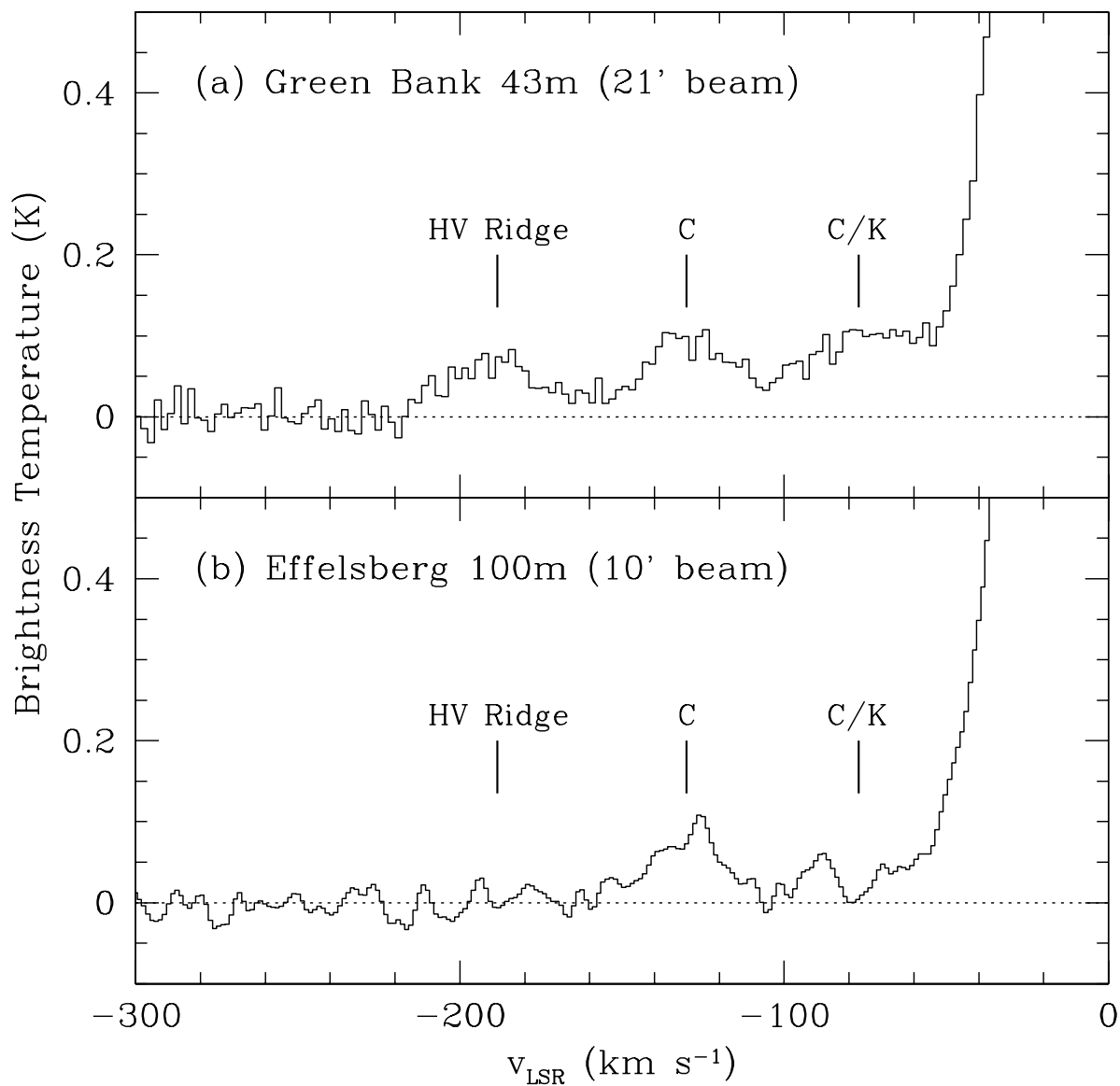}
\caption[]{\ion{H}{1} 21 cm emission profiles in the
direction of 3C 351, recorded with (a) the NRAO 43m
telescope (Murphy, Sembach, \& Lockman 2002), and (b) the
Effelsberg 100m telescope (Wakker et al. 2001). Brightness
temperature is plotted vs. LSR velocity. Complex C proper
and the high-velocity ridge are marked along with the
intermediate-velocity cloud Complex C/K. \ion{H}{1} column
densities derived from these data are listed in
Table~\ref{hi21}.\label{prof21}}
\end{figure}

A potentially serious source of systematic error in HVC abundance 
measurements is the large beam of the 21 cm observations usually used 
to estimate $N$(\ion{H}{1}). The large radio beam may dilute the HVC 21 
cm emission, or it may show emission from gas which is in the radio 
beam but is not present along the pencil beam to the UV source. 
Comparisons of $N$(\ion{H}{1}) measurements from different radio 
telescopes (with different beam sizes) indicate that the 10' Effelsberg 
beam is often sufficiently small to provide a reliable \ion{H}{1} 
column density (Wakker et al. 2001), but there are cases where even the 
Effelsberg beam is too large (e.g., Mrk 205, see Wakker 2001). 
Therefore it is worthwhile to review the available $N$(\ion{H}{1}) 
measurements for the 3C 351 HVCs. The HVC 21 cm emission profiles in 
the direction of 3C 351 observed with the NRAO 43m telescope (Murphy, 
Sembach, \& Lockman 2002) and the Effelsberg 100m telescope (Wakker et 
al. 2001) are shown in Figure~\ref{prof21}. \ion{H}{1} column densities 
in the 3C 351 HVCs derived from these data are summarized in 
Table~\ref{hi21} along with $N$(\ion{H}{1}) from the LDS data. The 
uncertainties listed in Table~\ref{hi21} are statistical errors only. 
Wakker et al. (2001) note that the systematic uncertainties in the 3C 
351 \ion{H}{1} column densities are likely much larger than the 
statistical errors: they estimate that systematic errors lead to an 
uncertainty of $\sim1.5 \times 10^{18}$ cm$^{-2}$ in $N$(\ion{H}{1}). 
With this uncertainty, the $N$(\ion{H}{1}) measurements in 
Table~\ref{hi21} for Complex C are in reasonable agreement. For Complex 
C, we adopt $N$(\ion{H}{1}) from the 
smallest-beam observation (Effelsberg), but with the larger systematic 
uncertainty reported by Wakker et al., i.e., $N$(\ion{H}{1}) = $(4.2 
\pm 1.5) \times 10^{18}$ cm$^{-2}$. However, while both Complex C 
proper and the 
high-velocity ridge are apparent in the NRAO and LDS data, only Complex 
C is clearly detected in the Effelsberg observation. This suggests that 
the observations with larger beams are picking up emission which is not 
present along the pencil beam to 3C 351. Consequently, we take the NRAO 
Green Bank $N$(\ion{H}{1}) as an {\it upper limit} for the 
high-velocity ridge, and we derive lower limits on the HV ridge 
abundances below. Similarly, the \ion{H}{1} column for Complex C/K is 
highly uncertain given the systematic uncertainty above, and we can 
only place lower limits on the metallicity of this IVC. However, these 
lower limits will turn out to be useful (\S \S 5.3,5.4).

\begin{deluxetable}{lcccccc}
\tablecolumns{7}
\tablewidth{0pc}
\tabletypesize{\footnotesize}
\tablecaption{Intermediate-Velocity and High-Velocity Cloud
\ion{H}{1} Column Densities toward 3C 351\label{hi21}}
\tablehead{ \ & \multicolumn{2}{c}{\underline{Effelsberg
100m (10$^\prime$ beam)\tablenotemark{a}}} &
\multicolumn{2}{c}{\underline{Green Bank 43m (21$^\prime$
beam)\tablenotemark{b}}} &
\multicolumn{2}{c}{\underline{Dwingeloo 25m (35$^\prime$
beam)\tablenotemark{c}}} \\
\ \ High-Velocity Cloud \ \ & $v_{\rm LSR}$ &
$N$(H~I)\tablenotemark{d} & $v_{\rm LSR}$ &
$N$(H~I)\tablenotemark{d} & $v_{\rm LSR}$ &
$N$(H~I)\tablenotemark{d} \\
 \ & (\kms ) & ($10^{18}$ cm$^{-2}$) & (\kms ) & ($10^{18}$
cm$^{-2}$) & (\kms ) & ($10^{18}$ cm$^{-2}$) }
\startdata
High-velocity ridge\dotfill & \nodata & \nodata & $-182.2$
& 4.19 & $-191$ & $4.5\pm 0.9$ \\
Complex C (proper)\dotfill & $-129$ & $4.2\pm 0.3$ &
$-128.2$ & 6.01 & $-126$ & $4.5\pm 0.6$ \\
Complex C/K\dotfill & $-89$ & $1.3\pm 0.2$ & $-83.0$ & 4.08
& $-69$ & $4.0\pm 0.5$
\enddata
\tablenotetext{a}{From Wakker et al. (2001).}
\tablenotetext{b}{From Lockman et al. (2002).}
\tablenotetext{c}{From Wakker et al. (2001), derived from
the Leiden-Dwingeloo Survey (Hartmann \& Burton 1997).}
\tablenotetext{d}{Listed column density uncertainties
include only statistical error, and uncertainties are not
reported by Lockman et al. (2002) for the Green Bank 43m
measurements. Wakker et al. (2001) discuss several sources
of systematic error, and they estimate that these lead to
an uncertainty of $\sim 1.5 \times 10^{18}$ cm$^{-2}$ in
$N$(\ion{H}{1}).}
\end{deluxetable}
\clearpage

\begin{figure}
\epsscale{1.0}
\plotone{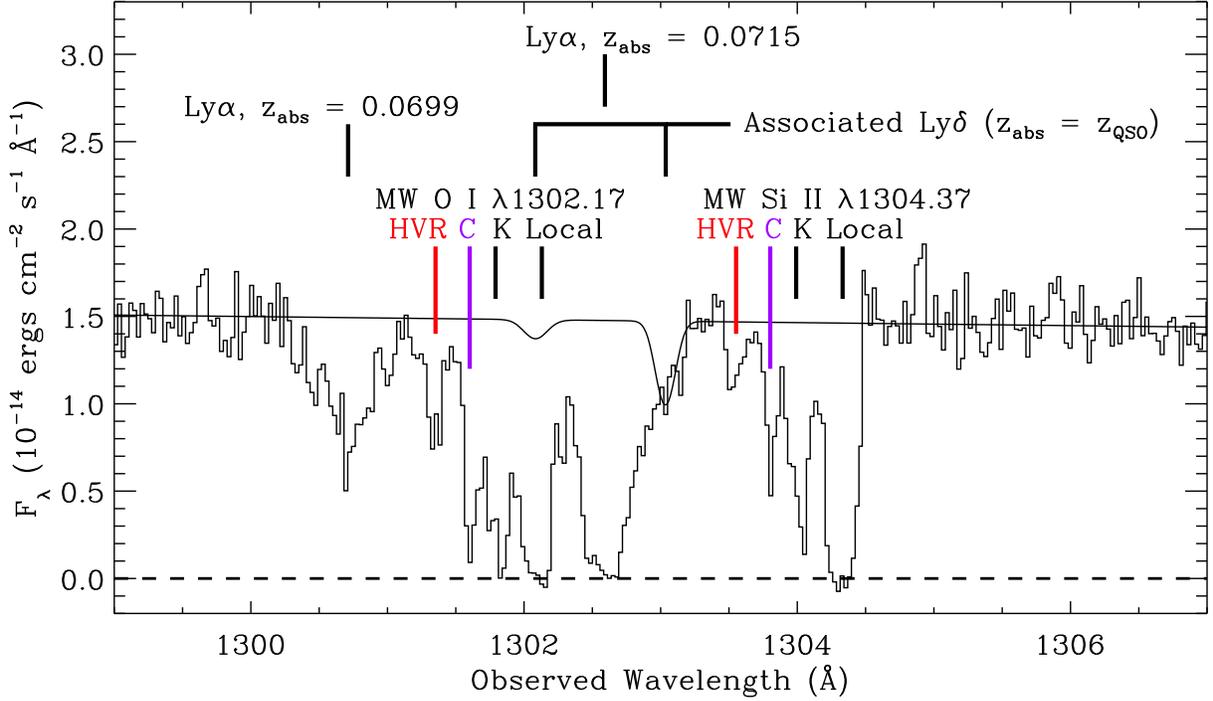}
\caption[]{Portion of the FUV spectrum of 3C 351 obtained
with STIS in the E140M echelle mode (FWHM $\approx$ 7 \kms
) plotted versus observed wavelength. For display purposes,
the spectrum has been binned (2 pixels into 1) in this
figure (all other figures, as well as the line
measurements, make use of the full resolution, unbinned
spectrum). This region of the spectrum shows the absorption
profiles of the \ion{O}{1} $\lambda$1302.2 and \ion{Si}{2}
$\lambda$1304.4 lines due to the ISM of the Milky Way as
well as several extragalactic absorption lines. Four main
components are readily apparent in the Galactic \ion{O}{1}
and \ion{Si}{2} profiles; these are indicated with tick
marks immediately above the spectrum. We are mainly
interested in the high-velocity absorption lines associated
with Complex C and the ``high-velocity ridge'' (HVR, see \S
2), which are marked with longer tick marks. Absorption
arising in the intermediate-velocity cloud Complex C/K and
local gas is also identified with short ticks. The thin
solid line indicates the fitted continuum with the
predicted Ly$\delta$ lines due to the associated absorption
systems at \zabs\ = 0.3709 and 0.3719 superimposed (see
Yuan et al. 2002). The other extragalactic lines in this
wavelength range are identified as \lya at \zabs\ = 0.0699
and 0.0715 (see text for further details).\label{oispec}}
\end{figure}

\begin{figure}
\plottwo{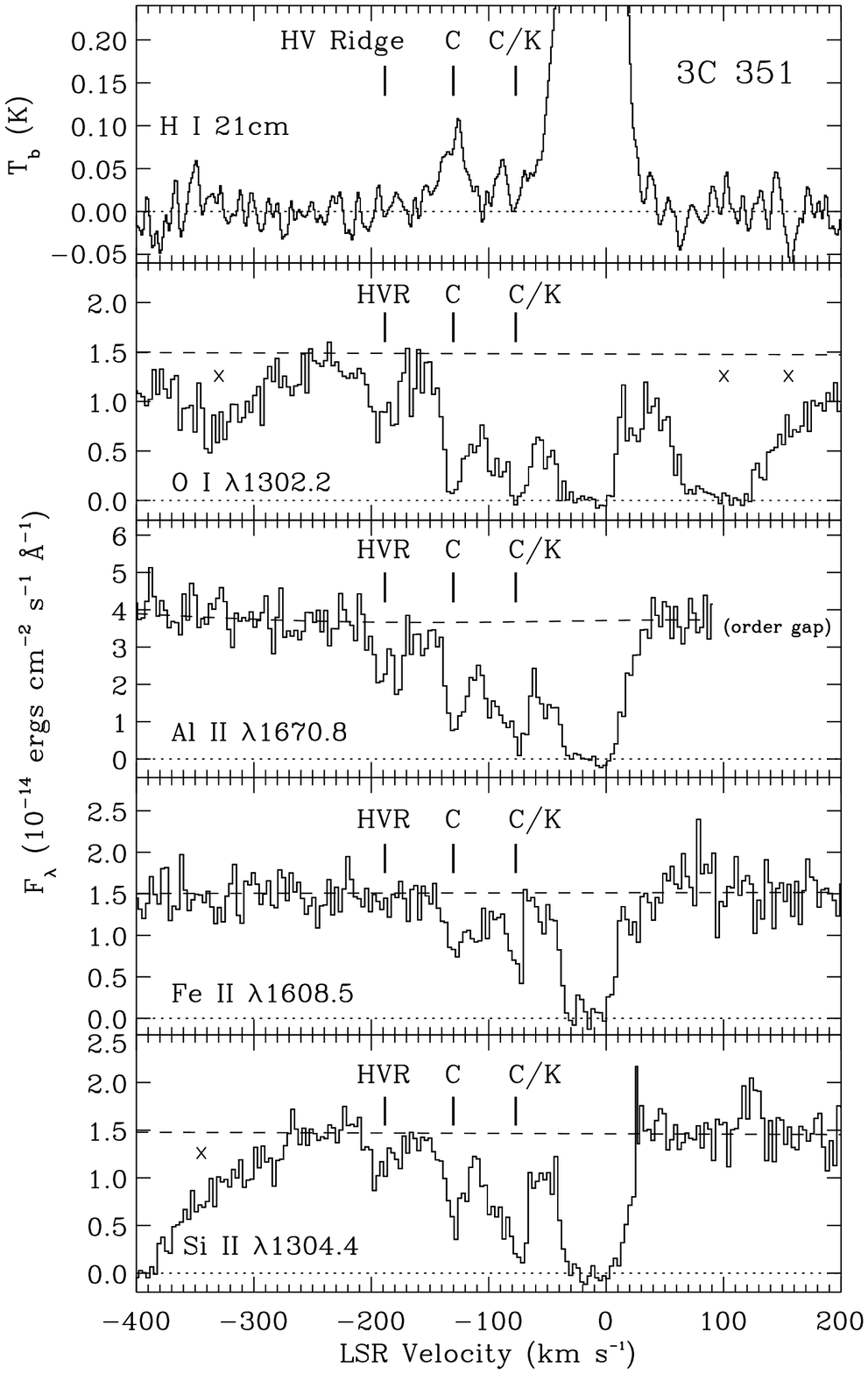}{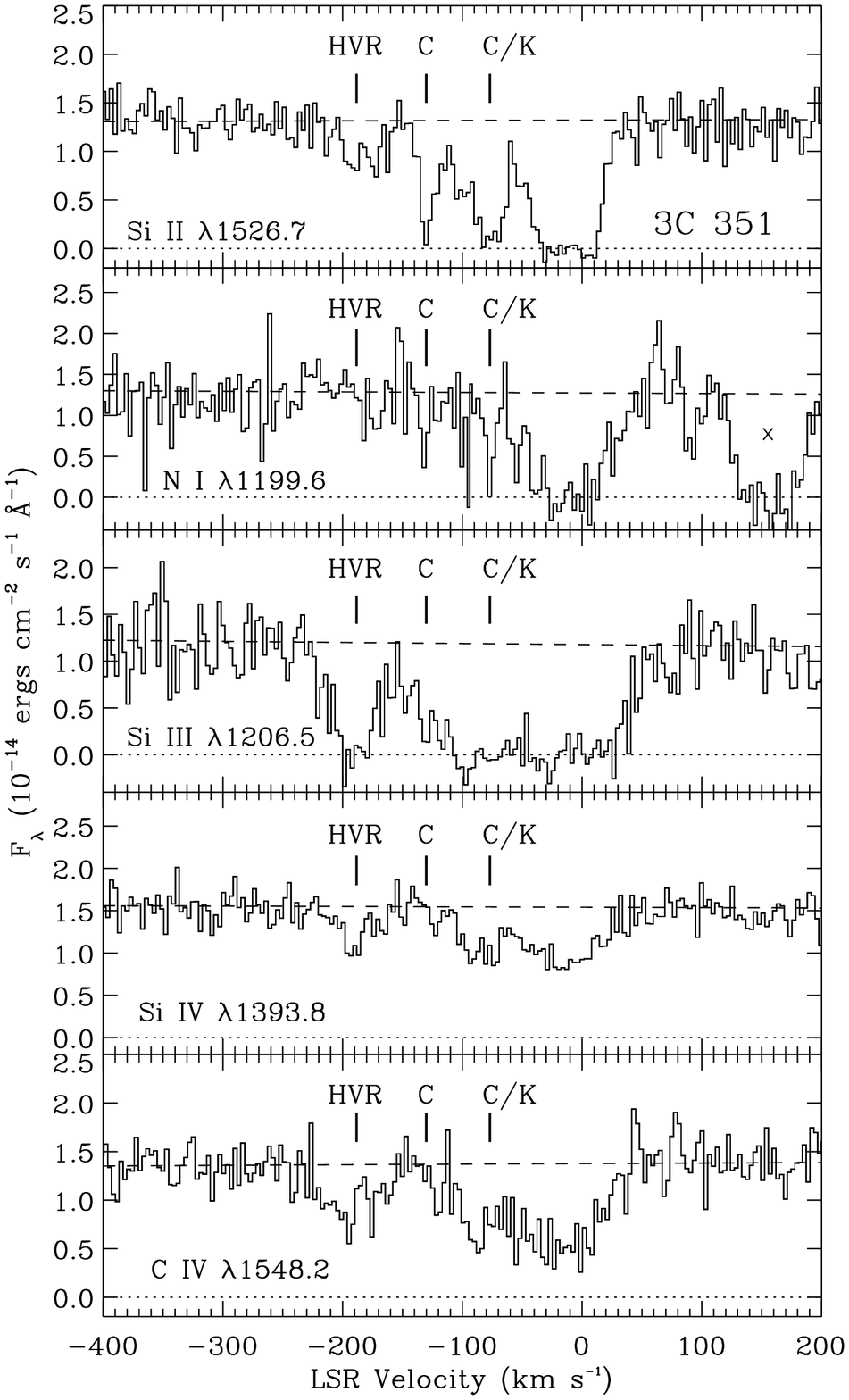}
\caption[]{Effelsberg 21 cm emission profile (top left
panel) and selected UV absorption line profiles of Milky
Way lines detected toward 3C 351, plotted versus Local
Standard-of-Rest (LSR) velocity. Brightness temperature is
plotted in the top left panel, and observed flux is shown
in the other panels. Dashed lines show the fitted continua
and dotted lines mark the flux zero level. Absorption lines
due to Complex C and the ``High-Velocity Ridge'' (HVR, see
text) are indicated with tick marks. Unrelated absorption
features are also marked ($\times$).\label{stack}}
\end{figure}
\clearpage

\section{STIS Echelle Spectroscopy\label{secobs}}

We now turn to the UV absorption line data. The STIS 
observations of 3C 351 are fully described in \S 2 of Yuan 
et al. (2002). Briefly, the QSO was observed with the E140M 
echelle mode with the $0\farcs 2 \times 0\farcs 06$ slit; 
this mode provides 7 \kms\ resolution (FWHM) and covers the 
1150 $-$ 1710 \AA\ range with only a few gaps between 
orders at the longest wavelengths (see Woodgate et al. 1998 
and Kimble et al. 1998 for further details). The data were 
reduced following standard procedures with software 
developed by the STIS Investigation Definition Team. 
Scattered light was removed using the procedure developed 
by the STIS Team (Bowers et al. 1998), which corrects for 
echelle scatter as well as other sources of scattered 
light.

A sample of the final spectrum is shown in 
Figure~\ref{oispec}. This somewhat complicated portion of 
the spectrum shows the absorption profiles of the important 
Milky Way \ion{O}{1} $\lambda$1302.2 and \ion{Si}{2} 
$\lambda$1304.4 profiles as well as several extragalactic 
absorption lines. Absorption profiles of several species of 
interest are plotted versus LSR velocity in 
Figure~\ref{stack} along with the \ion{H}{1} 21 cm emission 
observed in the direction of 3C 351 by Wakker et al. (2001) 
with the 100~m Effelsberg telescope. The UV absorption 
lines show four distinct components. The two 
highest-velocity components at \vlsr\ = $-128$ and $-190$ 
\kms\ are close to the observed 21 cm velocities of Complex 
C and the high-velocity ridge. Similarly, the UV lines at 
\vlsr\ $\approx -82$ \kms\ are close to the velocity of IVC 
C/K. Analysis of the Complex C/K absorption lines is 
complicated by substantial blending with adjacent 
components. Nevertheless, the data indicate that IVC C/K 
has a higher metallicity than HVC Complex C (\S 5.4).

In this paper, we are primarily interested in Complex C and 
the high-velocity ridge. However, absorption from other 
Galactic and extragalactic clouds can introduce systematic 
errors that must be considered. For example, 3C 351 has a 
dramatic ``associated'' absorption system (i.e., $z_{\rm 
abs} \approx z_{\rm QSO}$) that affects the spectrum near 
the \ion{N}{5}, \ion{O}{6}, and Lyman series lines at the 
redshift of the QSO (Yuan et al. 2002, and references 
therein). The associated Ly$\delta$ lines at \zabs\ = 
0.3709 and 0.3719 fall near the Galactic \ion{O}{1} 
$\lambda 1302.2$ profile, which is a crucial profile for 
our analysis. We can predict the strength of these 
contaminating Ly$\delta$ lines based on fits to the other 
Lyman series lines in the STIS spectrum from Yuan et al. 
(2002). The thin solid line in Figure~\ref{oispec} shows 
the predicted Ly$\delta$ lines superimposed on the fitted 
continuum. Fortunately, as can be seen from 
Figure~\ref{oispec}, these associated Ly$\delta$ lines have 
no impact on the high-velocity \ion{O}{1} lines of primary 
interest. The nearby Ly$\alpha$ lines at \zabs\ = 0.0715 
and particularly 0.0699 could be a more serious source of 
contamination than the associated Ly$\delta$ lines. These 
Ly$\alpha$ lines appear to be mostly separated from the 
Milky Way \ion{O}{1} and \ion{Si}{2} profiles, but since 
\lya lines are clustered on velocity scales of a few 
hundred \kms\ (Penton et al. 2000), it remains possible 
that unseen \lya lines clustered around the strong \zabs\ = 
0.0699 and 0.0715 absorbers affect the Galactic \ion{O}{1} 
and/or \ion{Si}{2} profiles in Figure~\ref{oispec}. 
However, a detailed  comparison of the Milky Way profiles 
in the following section indicates that there is little or 
no contamination present in the \ion{O}{1} and \ion{Si}{2} 
lines.

The Galactic \ion{Si}{2} $\lambda 1193.3$ transition is 
badly blended with \ion{C}{3} $\lambda 977.0$ absorption 
from a Lyman limit system at \zabs\ = 0.2210, so we do not 
use this transition for the HVC analysis. Also, the Milky 
Way \ion{Si}{2} $\lambda 1190.4$ and \ion{S}{3} $\lambda 
1190.2$ transitions are highly blended, and we do not trust 
information from either of these lines. Finally, we note 
that the Galactic \ion{Si}{2} $\lambda 1260.4$ and 
\ion{C}{2} $\lambda 1334.5$ transitions are severely 
saturated and consequently do not provide useful 
constraints.

\section{Absorption Line Measurements\label{secmeas}}

To measure the line column densities and to assess the 
impact of unresolved saturation on the absorption lines of 
interest, we mainly rely on the apparent column density 
technique (Savage \& Sembach 1991), but some supporting 
measurements are also derived from Voigt profile fitting, 
using the software of Fitzpatrick \& Spitzer (1997) with 
the line spread functions from the STIS Instrument Handbook 
(Leitherer et al. 2001). The apparent column density is 
constructed using the following expression,
\begin{equation}
N_{\rm a}(v) \ = \ (m_{\rm e}c/\pi e^{2})(f\lambda )^{-1}\tau _{\rm 
a}(v) \ 
 = \ 3.768 \times 10^{14} (f\lambda )^{-1} {\rm ln}[I_{\rm c}(v)/I(v)].
\end{equation}
This provides the instrumentally broadened column density per unit 
velocity [in atoms cm$^{-2}$ (\kms )$^{-1}$], where $f$ is the 
transition oscillator strength, $\lambda$ is the transition wavelength 
(the numerical coefficient is for $\lambda$ in \AA ), $I(v)$ is the 
observed line intensity and $I_{\rm c} (v)$ is the estimated continuum 
intensity at velocity $v$. By comparing two or more resonance lines of 
a given species with adequately different $f\lambda$ values, the 
profiles can be checked for unresolved saturation and a correction can, 
in some cases, be applied (Jenkins 1996). If the \nav\ profiles of the 
resonance lines are in good agreement, then the lines are not affected 
by unresolved saturation, and the profiles can be integrated to obtain 
the total column, $N_{\rm tot} = \int N_{\rm a} (v) dv.$

High-velocity absorption line equivalent widths and apparent column 
densities, integrated across the velocity ranges that delimit Complex C 
proper and the high-velocity ridge, are summarized in 
Table~\ref{lineprop}. These measurements were made using the methods of 
Sembach \& Savage (1992) and include continuum placement uncertainty 
and a 2\% flux zero point uncertainty in addition to the statistical 
uncertainties. Column densites of well-detected species determined from 
Voigt profile fitting ($N_{\rm pf}$) are also listed. Equivalent widths 
and column densities for the 
intermediate-velocity C/K cloud are provided in Table~\ref{ivcprop}. 
The abundance measurements presented in the last two columns of 
Tables~\ref{lineprop}-\ref{ivcprop} are discussed in \S \ref{secabun}.

\begin{deluxetable}{rcclccllll}
\rotate
\tabletypesize{\footnotesize}
\tablewidth{0pc}
\tablecaption{High-Velocity Absorption Lines toward 3C 351: 
Complex C and the High-Velocity Ridge\label{lineprop}}
\tablehead{\ \ Species \ \ & $\lambda 
_{0}$\tablenotemark{a} & $f$\tablenotemark{a} & \ \ Cloud & 
$<v>$\tablenotemark{b} & $W_{\lambda}$\tablenotemark{c} & \ 
\ log $N_{\rm a}$\tablenotemark{c} & \ \ log $N_{\rm 
pf}$\tablenotemark{d} & \ \ 
[X$^{i}$/H~I]\tablenotemark{e} & \ \ 
[X/H]\tablenotemark{f} \\
 \ & (\AA ) & \ & \ & (km s$^{-1}$) & (m\AA ) & \ & \ & \ & 
\ }
\startdata
O~I\dotfill & 1302.17 & 0.0519 & Complex C & $-
128\pm 1$ & 115.1$\pm$4.9 & 14.45$\pm$0.04 & 
14.63$^{+0.10}_{-0.14}$ & $-0.73^{+0.23}_{-0.21}$ & $-
0.76^{+0.23}_{-0.21}$ \\
    \             &   \     &   \    & HV Ridge & $-190\pm 
2$ & 81.8$\pm$8.1 & 14.12$\pm$0.05 & 14.17$\pm$0.04 & $> -
1.24$ & $> -1.26$ \\
\\
N~I\dotfill & 1199.55 & 0.130  & Complex C & $-
130\pm 3$ & (24.6$\pm$10.8) & $<$13.44\tablenotemark{g} & 
\nodata & $< -1.11$ & $< -1.37$ \\
    \             &   \     &    \   & HV Ridge & \nodata & 
(3.7$\pm$14.6) & $<$13.55\tablenotemark{g} & \nodata & 
\nodata & \nodata \\
\\
Si~II\tablenotemark{h}\dotfill  & 1304.37 & 0.0917 & 
Complex C & $-128\pm 1$ & 63.3$\pm$4.6 & 13.81$\pm$0.05 & 
13.77$\pm$0.04 & $-0.40 ^{+0.20}_{-0.14}$ & $-0.73 
^{+0.20}_{-0.14}$ \\
      \           &   \     &    \   & HV Ridge & $-183\pm 
5$ & 30.3$\pm$7.1 & 13.42$^{+0.08}_{-0.12}$ & 
13.50$\pm$0.05 & $> -0.70$ & $> -1.26$ \\
      \           & 1526.71 & 0.132  & Complex C & $-128\pm 
1$ & 86.1$\pm$6.5 & 13.72$\pm$0.05 & 13.77$\pm$0.04 & $-
0.40 ^{+0.20}_{-0.14}$ & $-0.73 ^{+0.20}_{-0.14}$ \\
      \           &    \    &   \    & HV Ridge & $-184\pm 
3$ & 68.8$\pm$10.5 & 13.49$^{+0.06}_{-0.08}$ & 
13.50$\pm$0.05 & $> -0.70$ & $> -1.26$ \\
\\
Al~II\dotfill  & 1670.79 & 1.83 & Complex C & $-
127\pm 1$ & 105.0$\pm$6.9 & 12.57$\pm$0.04 & 12.56$\pm$0.05 
& $-0.52 ^{+0.21}_{-0.15}$ & $-0.88 ^{+0.21}_{-0.15}$ \\
      \           &  \      &  \   & HV Ridge & $-182\pm 3$ 
&  70.1$\pm$11.6 & 12.30$^{+0.06}_{-0.08}$ & 12.33$\pm$0.05 
& $> -0.79$ & $> -1.36$ \\
\\
S~II\dotfill  & 1259.52 & 0.0166 & Complex C & $-
124\pm 6$ & (12.0$\pm$5.6) & $\leq 14.0$\tablenotemark{g} & 
\nodata & $< 0.05$ & $< -0.31$ \\
     \            &    \    &    \   & HV Ridge & \nodata & 
(1.1$\pm$7.9) & $\leq 14.1$\tablenotemark{g} & \nodata & 
\nodata & \nodata \\
Fe~II\dotfill  & 1608.45 & 0.0580 & Complex C & $-
125\pm 2$ & 63.1$\pm$8.7 & 13.78$^{+0.06}_{-0.08}$ & 
13.88$^{+0.14}_{-0.21}$ & $-0.29^{+0.22}_{-0.17}$ & $-
0.54^{+0.22}_{-0.17}$ \\
    \             &    \    &   \    & HV Ridge & \nodata & 
(12.4$\pm$13.8) & $\leq 13.6$\tablenotemark{g} & \nodata & 
\nodata & \nodata \\
\\
Si~III\dotfill  & 1206.50 & 1.67   & Complex C & $-127 
\pm 2$\tablenotemark{g} & 115.2$\pm$7.6\tablenotemark{i} & 
12.91$\pm$0.06\tablenotemark{i} & \nodata & \nodata & 
\nodata \\
    \             &   \     &  \     & HV Ridge & $-189 \pm 
3$ & 196.0$\pm$12.5 & $> 13.07$\tablenotemark{j} & 
13.40$\pm$0.10\tablenotemark{j} & \nodata & \nodata 
\tablebreak \\
\\
Si~IV\dotfill  & 1393.76 & 0.514  & Complex C & $-
118\pm 13$ & (7.1$\pm$6.2) & $\leq 12.4$\tablenotemark{g} & 
\nodata & \nodata & \nodata \\
    \             &    \    &  \     & HV Ridge & $-188\pm 
4$ & 45.1$\pm$8.4 & 12.77$^{+0.08}_{-0.09}$ & 
12.79$\pm$0.07 & \nodata & \nodata \\
\\
C~IV\dotfill  & 1548.20 & 0.191  & Complex C & $-
120\pm 8$ & (14.6$\pm$8.9) & $\leq 13.0$\tablenotemark{g} & 
\nodata & \nodata & \nodata \\
    \             &    \    &  \     & HV Ridge & $-191\pm 
3$ & 100.2$\pm$11.6 & 13.50$\pm$0.06 & 13.54$^{+0.16}_{-
0.26}$ & \nodata & \nodata
\\
\\
\enddata
\tablenotetext{a}{Rest-frame vacuum wavelength and 
oscillator strength from Morton (2002) or Morton (1991).}
\tablenotetext{b}{Profile-weighted mean velocity of the 
line, $<v> = \int v[1 - I(v)/I_{\rm c}(v)] dv/\int [1 - 
I(v)/I_{\rm c}(v)] dv.$}
\tablenotetext{c}{Integrated equivalent width and apparent 
column density (in cm$^{-2}$). For the High-Velocity Ridge, 
all quantities are integrated from $v_{\rm LSR} = -230$ to 
$-155$ \kms , and the Complex C absorption lines are 
integrated over $-150 \leq v_{\rm LSR} \leq -110$. 
Equivalent widths in parentheses have less than 3$\sigma$ 
significance, and we do not consider these lines to be 
reliably detected.}
\tablenotetext{d}{Column density (in cm$^{-2}$) from Voigt 
profile fitting.}
\tablenotetext{e}{Implied logarithmic abundance {\it if 
ionization corrections are neglected}, i.e., 
[X$^{i}$/H~I] = log $N$(X$^{i}$)/$N$(H~I) - 
log (X/H)$_{\odot}$.}
\tablenotetext{f}{\scriptsize Logarithmic abundance obtained by 
applying the ionization correction from the best-fitting 
models presented in the text. In this table we have used 
the collisional ionization equilibrium corrections, but 
very similar results are obtained from the CLOUDY model 
(see\S \S~\ref{seccollion},\ref{secphot},\ref{sechvr}). 
Error bars include column density uncertainties and solar 
reference abundance uncertainties but do not reflect 
uncertainties in the ionization correction.}
\tablenotetext{g}{4$\sigma$ upper limit assuming the linear 
curve-of-growth applies.}
\tablenotetext{h}{The Voigt profile fitting column is from 
a simultaneous fit to the 1304.37 and 1526.71 \AA\ lines. 
Similarly we combined the information from these two 
transitions for the directly integrated column. For the 
final integrated Si~II column density, we adopt the 
following weighted averages of the 1304.37 and 1526.71 \AA\ 
measurements. Complex C: log $N$(Si~II) = 
13.76$\pm$0.03; high-velocity ridge: log $N$(Si~II) = 
13.46$\pm$0.06.}
\tablenotetext{i}{Formal error bars may underestimate the 
true uncertainty due to strong blending with lower velocity 
gas (see Figure~\ref{stack}).}
\tablenotetext{j}{Strongly saturated absorption.}
\end{deluxetable}

\begin{deluxetable}{rcccllll}
\tabletypesize{\footnotesize}
\tablewidth{0pc}
\tablecaption{Intermediate-Velocity Absorption Lines toward 
3C 351: Complex C/K\tablenotemark{a}\label{ivcprop}}
\tablehead{\ \ Species \ \ & $\lambda _{0}$ & $<v>$ & 
$W_{\lambda}$ & \ \ log $N_{\rm a}$ & \ \ log $N_{\rm pf}$ 
& \ \ [X$^{i}$/H~I]\tablenotemark{b} & \ \ 
[X/H]\tablenotemark{c} \\
 \ & (\AA ) & (km s$^{-1}$) & (m\AA ) & \ & \ & \ & \ }
\startdata
O~I\dotfill & 1302.17 & $-82\pm 1$ & 
180.6$\pm$5.4\tablenotemark{d} & $>14.66$\tablenotemark{e} 
& 14.85$^{+0.18}_{-0.15}$\tablenotemark{d} & $\geq -0.50$ & 
$\geq -0.52$ \\
N~I\dotfill & 1199.55 & $-83\pm 10$ & 66.5$\pm$13.5 
& 13.63$^{+0.10}_{-0.14}$ & \nodata & $\geq -0.91$ & $\geq 
-0.69$ \\
Si~II\dotfill & 1304.37 & $-81\pm 1$ & 131.3$\pm$4.9 
& $>14.23$\tablenotemark{e} & 14.35$^{+0.13}_{-
0.10}$\tablenotemark{f} & $\geq 0.20$ & $\geq -0.52$ \\
   \                & 1526.71 & $-81\pm 1$ & 181.1$\pm$7.0 
& $>14.09$\tablenotemark{e} & 14.35$^{+0.13}_{-
0.10}$\tablenotemark{f} & $\geq 0.20$ & $\geq -0.52$ \\
Al~II\dotfill & 1670.79 & $-82\pm 1$ & 
191.1$\pm$7.2\tablenotemark{d} & 
12.88$\pm$0.04\tablenotemark{d,g} & $\sim 
13.1$\tablenotemark{h} & $\geq -0.20$ & $\geq -0.88$ \\
S~II\dotfill & 1259.52 & $-86\pm 4$ & 26.1$\pm$6.2 & 
14.10$^{+0.10}_{-0.12}$ & \nodata & $\geq 0.16$ & $\geq -
0.46$ \\
Fe~II\dotfill & 1608.45 & $-84\pm 2$ & 79.9$\pm$10.2 
& 13.91$\pm$0.07 & 13.86$\pm 0.10$ & $\geq -0.15$ & $\geq -
0.64$ \\
Si~IV\dotfill & 1393.76 & $-83\pm 2$ & 68.9$\pm$6.3 & 
12.98$\pm$0.05 & 13.03$\pm 0.06$ & \nodata & \nodata \\
C~IV\dotfill & 1548.20 & $-82\pm 1$ & 
110.6$\pm$9.0\tablenotemark{b} & 
13.58$\pm$0.05\tablenotemark{b} & 13.29$\pm 
0.07$\tablenotemark{b} & \nodata & \nodata
\enddata
\tablenotetext{a}{See Table~\ref{lineprop} for definitions 
of the quantities in this table. The equivalent widths and 
apparent column densities are integrated over the velocity 
range of the intermediate-velocity C/K cloud, from $v_{\rm 
LSR}$ = $-110$ to $-59$ km s$^{-1}$.}
\tablenotetext{b}{Implied logarithmic abundance {\it if 
ionization corrections are neglected}, i.e., 
[X$^{i}$/H~I] = log $N$(X$^{i}$)/$N$(H~I) - 
log (X/H)$_{\odot}$.}
\tablenotetext{c}{Logarithmic abundance obtained by 
applying the ionization correction from the best-fitting 
CLOUDY model presented in the text (\S~\ref{secivc}). Error 
bars include column density uncertainties and solar 
reference abundance uncertainties but do not reflect 
uncertainties in the ionization correction.}
\tablenotetext{d}{Line is strongly blended with adjacent 
absorption features.}
\tablenotetext{e}{Saturated absorption line.}
\tablenotetext{f}{Simultaneous fit to the Si~II 
$\lambda$1304.37 and $\lambda$1526.71 lines.}
\tablenotetext{g}{Apparent column density may underestimate 
the true column due to unresolved saturation.}
\tablenotetext{h}{Poorly constrained result due to strong 
blending.}
\end{deluxetable}
\clearpage

As noted in \S 1, the most important species for metallicity 
measurements is \ion{O}{1}. The only \ion{O}{1} transition available in 
the direction of 3C 351 is the 1302.2 \AA\ line.\footnote{In principle, 
weaker O~I transitions can be observed with {\it FUSE} at 915 $< 
\lambda <$ 1050 \AA . Unfortunately, there is not sufficient flux from 
the QSO in this wavelength range for absorption line measurements 
because the spectrum is strongly attenuated below 1115 \AA\ by an 
optically thick Lyman limit absorber at \zabs\ = 0.2210 (see Mallouris 
et al., in preparation).} In the 
high-velocity ridge, $N$(\ion{O}{1}) appears to be well-constrained 
from the 1302.2 \AA\  profile. In Complex C, on the other hand, the 
1302.2 \AA\ line does not go to zero flux in the core, but it is quite 
strong (see Figs.~\ref{oispec}-\ref{stack}). Consequently, it is 
important to consider whether the Complex C \ion{O}{1} column might be 
underestimated due to unresolved saturation. We believe that the 
\ion{O}{1} line is {\it not} seriously saturated based on the following 
evidence. We first note that the \ion{Si}{2} lines show no indications 
of unresolved saturation. Figure~\ref{navprofs}a compares the \nav\ 
profiles of the 1304.4 and 1526.7 \AA\ lines of Galactic \ion{Si}{2}, 
transitions that differ in $f\lambda$ by 0.23 dex. These \ion{Si}{2} 
\nav\ profiles are in good agreement,\footnote{Unresolved saturation 
would be manifested by a higher apparent column in the weaker 
transition (1304.4) compared to the stronger transition. The highest 
outlying point in the \nav\ profiles in Figure~\ref{navprofs}a is from 
the 1526.7 \AA\ transition, the opposite of the expected effect.} which 
indicates that the lines are not significantly saturated. If \ion{O}{1} 
is affected by unresolved saturation, then the \ion{O}{1} \nav\ profile 
should differ from the \ion{Si}{2} \nav\ profiles due to the 
underestimation of $N$(\ion{O}{1}) in the saturated pixels. 
Figure~\ref{navprofs}b compares the \ion{O}{1} apparent column density 
to a weighted composite \ion{Si}{2} \nav\ profile constructed from the 
1304.4 and 1526.7 profiles following the procedure of Jenkins \& 
Peimbert (1997), with the \ion{O}{1} profile scaled down by a factor of 
five. From Figure~\ref{navprofs}b, we see that the \ion{O}{1} and 
\ion{Si}{2} profiles have very similar shapes. This suggests that there 
is little unresolved saturation within the \ion{O}{1} profile. We 
notice from Table~\ref{lineprop} that Voigt profile fitting gives a 
somewhat higher $N$(\ion{O}{1}) than direct \nav\ integration, although 
the \ion{O}{1} columns from the two methods agree within the 1$\sigma$ 
uncertainties. This could reflect a small amount of saturation. To be 
conservative, we adopt the higher $N$(\ion{O}{1}) value for our 
subsequent analysis. However, we will consider the lower \ion{O}{1} 
column in our analysis, and we will find that our conclusions are 
essentially the same.

The comparison of the \ion{O}{1} and \ion{Si}{2} \nav\ profiles in 
Figure~\ref{navprofs}b also indicates that there is little 
contamination of the \ion{O}{1} lines from the nearby \lya absorbers 
discussed in \S 3 and shown in Figure~\ref{oispec}. Significant optical 
depth from such contamination would cause the \ion{O}{1} and 
\ion{Si}{2} profiles to have different shapes at velocities where the 
contamination occurs. The \ion{O}{1} and \ion{Si}{2} profiles have 
similar velocity structure, and moreover the \ion{O}{1}/\ion{Si}{2} 
column density ratio is very similar in Complex C and the high-velocity 
ridge. The \ion{Al}{2} and \ion{Si}{2} \nav\ profiles are also 
remarkably alike, as shown in Figure~\ref{navprofs}d. The \ion{Fe}{2} 
\nav\ profile is compared to the composite \ion{Si}{2} profile in 
Figure~\ref{navprofs}c. \ion{Fe}{2} is clearly detected in Complex C, 
but the substantially greater noise limits comparisons of C and the HV 
ridge for this species. The similarity of the \ion{O}{1}, \ion{Si}{2}, 
and \ion{Al}{2} profiles in Figure~\ref{navprofs} would seem to suggest 
that the physical conditions and relative abundances in Complex C and 
the high-velocity ridge are roughly the same. However, the behavior of 
the higher ionization stages (\ion{Si}{4} and \ion{C}{4}) is entirely 
different in Complex C and the high-velocity ridge, as shown in in 
Figure~\ref{navprofs}e-f. Very little high-ion absorption is apparent 
in the velocity range of Complex C, but \ion{Si}{4} and \ion{C}{4} are 
clearly detected in the high-velocity ridge, with component structure 
similar to that of the lower ionization stages. The \ion{Si}{4} and 
\ion{C}{4} profiles therefore indicate that there is a important 
difference in the physical conditions of Complex C and the HV ridge. We 
discuss this issue further in the following section.

\begin{figure}
\epsscale{0.9}
\plotone{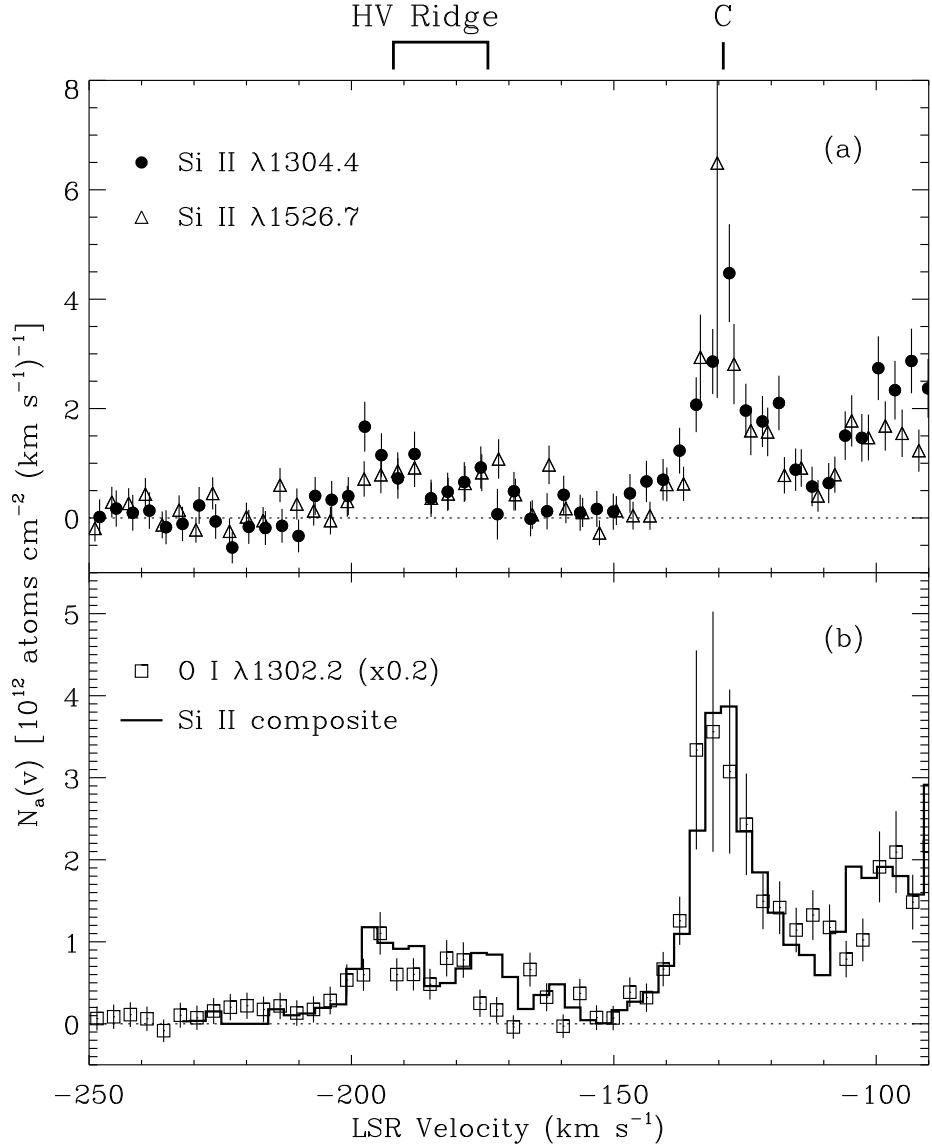}
\caption[]{Apparent column density as a
function of LSR velocity (\nav , see text \S~\ref{secmeas})
of selected absorption lines detected in the high-velocity
clouds in the direction of 3C 351. (a) \nav\ profiles of
the \ion{Si}{2} transitions at 1304.4 \AA\ (filled circles)
and 1526.7 \AA\ (open circles). The good agreement of these
transitions indicates that the lines are not strongly
affected by unresolved saturation. (b) \ion{O}{1}
$\lambda$1302.2 profile scaled by 0.2 (open squares) vs. a
weighted composite \ion{Si}{2} profile (histogram). (c)
\ion{Fe}{2} $\lambda$1608.5 profile (filled circles) vs.
the composite \ion{Si}{2} profile (histogram). (d)
\ion{Al}{2} $\lambda$1670.8 profiles scaled by 17.0 (open
squares) vs. the composite \ion{Si}{2} profile (histogram).
(e) \ion{Si}{4} $\lambda$1393.8 profile scaled by 5.0
(filled circles) vs. the composite \ion{Si}{2} profile
(histogram). (f) \ion{C}{4} $\lambda$1548.2 profile (open
squares) vs. the composite \ion{Si}{2} profile
(histogram).\label{navprofs}}
\end{figure}

\begin{figure}
\figurenum{8}
\plotone{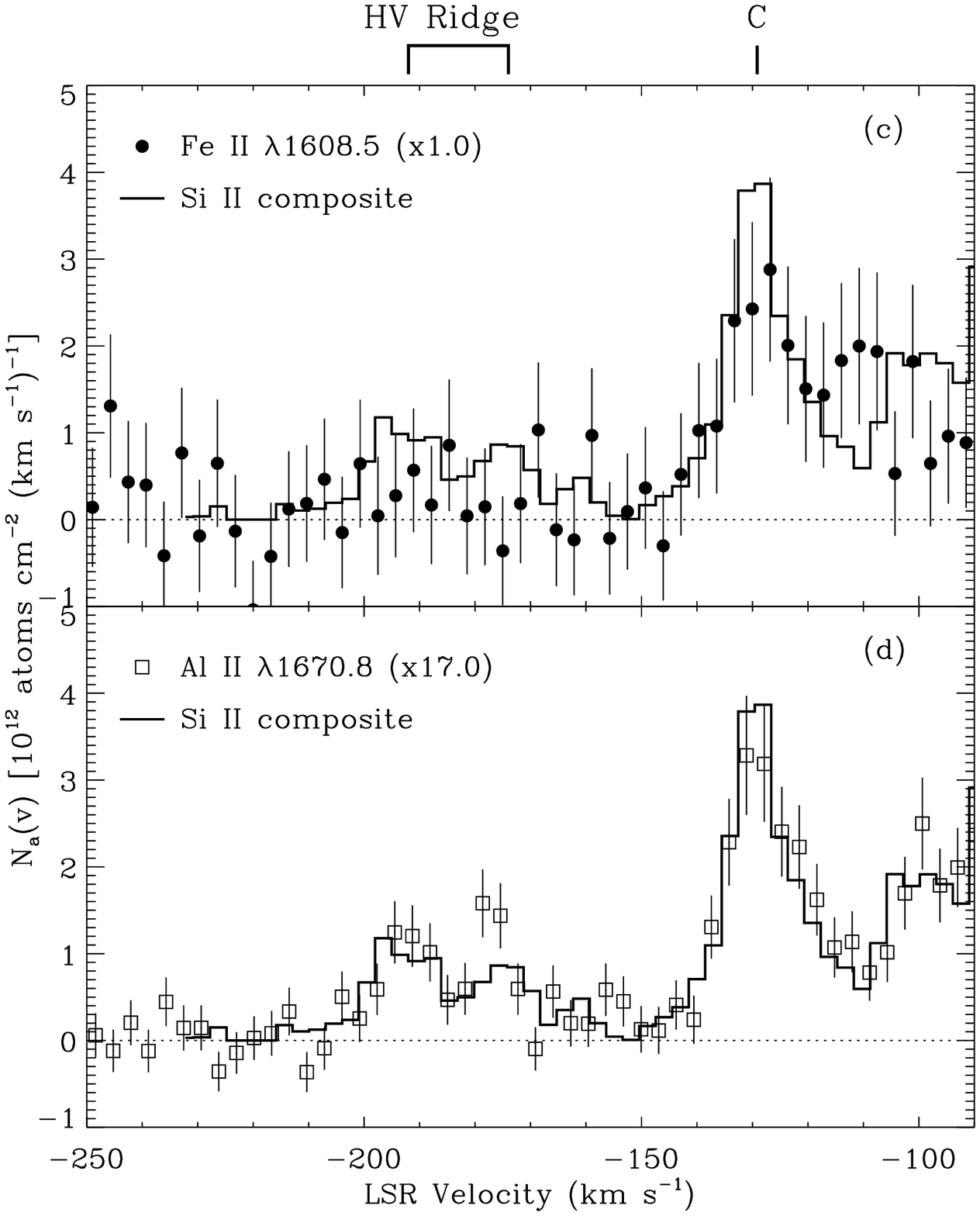}
\caption[]{continued}
\end{figure}

\begin{figure}
\figurenum{8}
\plotone{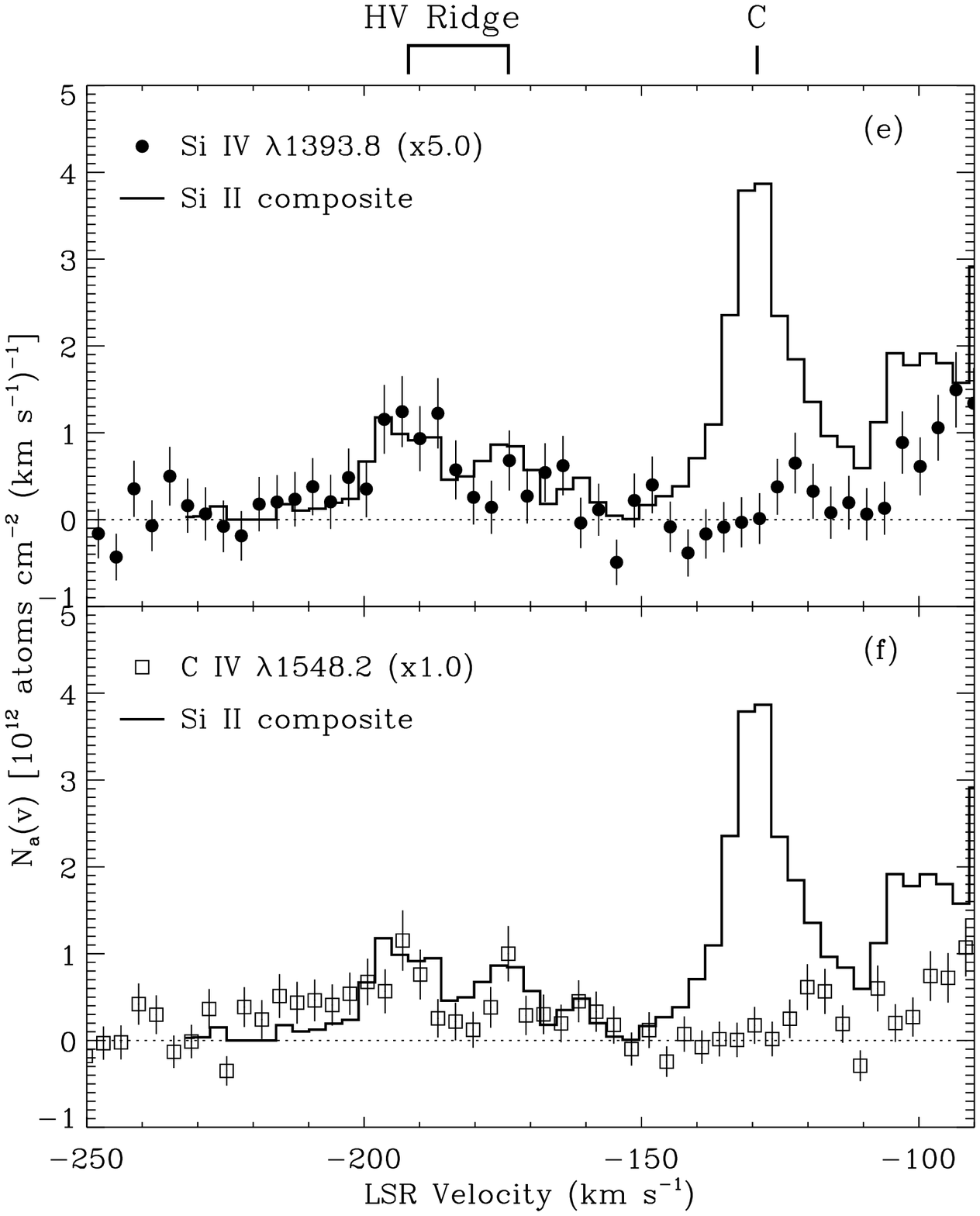}
\caption[]{continued}
\end{figure}
\clearpage

\section{Abundances and Ionization\label{secabun}}

We next consider the implications of the measurements presented above. 
We first argue that ionization corrections are important in the 3C 351 
HVCs (\S5.1). We then investigate the absolute and relative abundances 
in Complex C and the 
high-velocity ridge with the aid of ionization models (\S \S5.2$-$5.4).

\subsection{Preliminary Remarks}

In general, the logarithmic abundance of species X with respect to 
species Y is
\begin{equation}
\left[ \frac{\rm X}{\rm Y}\right] = {\rm log}\left( 
\frac{N({\rm X}^{i})}{N({\rm Y}^{i})}\right) + {\rm 
log}\left( \frac{f({\rm Y}^{i})}{f({\rm X}^{i})}\right) - 
{\rm log}\left( \frac{\rm X}{\rm Y}\right) _{\odot}, 
\label{metlim}
\end{equation}
where $N({\rm X}^{i})$ and $f({\rm X}^{i})$ are the column density and
ion fraction of the $i^{\rm th}$ ionization stage of species X (and
likewise for Y), and (X/Y)$_{\odot}$ is the solar reference
abundance.\footnote{We adopt the solar abundances reported by Holweger
  (2002) for the most abundant elements. The oxygen abundance from
  Holweger is in excellent agreement with the independent solar oxygen
  measurement reported by Allende Prieto, Lambert, \& Asplund (2001),
  and these solar abundances are close to the interstellar oxygen
  measurements in the vicinity of the Sun (Sofia \& Meyer 2001). Solar
  abundances of sulfur and aluminum are taken from Grevesse, Noels, \&
  Sauval (1996). } In many situations, the ionization correction
$f({\rm Y}^{i})/f({\rm X}^{i})$ can be neglected, e.g., if $ f({\rm
  Y}^{i}) = f({\rm X}^{i}) = 1$.  However, when we state that the
``ionization correction is decreasing'', this does not indicate
that the ionization correction is becoming negligible; this only
indicates that $f({\rm Y}^{i})/f({\rm X}^{i})$ is decreasing. In this
paper, to distinguish between measurements that have been corrected
for ionization and those which have not, we adopt the following
notation: uncorrected abundances are indicated by the observed ion,
e.g., [\ion{Si}{2}/\ion{H}{1}], while measurements which have had an
ionization correction applied are indicated without reference to a
particular ion, e.g., [Si/H]. The penultimate columns of
Tables~\ref{lineprop}-\ref{outerarmprop} list the uncorrected
abundances implied by our column density measurements. The last
columns of these tables summarize the ionization-corrected abundances,
using the ionization models described in the sections below. For the
high-velocity ridge and IVC C/K, we list lower limits on the
abundances because we only have upper limits on $N$(\ion{H}{1}) for
these components (see \S~\ref{sec21cm}). Ionization corrections can be
a large source of uncertainty; we present examples below. The
$N$(\ion{H}{1}) measurements are often the other main source of
uncertainty. We include our best estimates of the $N$(\ion{H}{1})
uncertainty in our abundance error bars. However, systematic errors in
$N$(\ion{H}{1}) due to radio beam effects can be much larger than
statistical errors, and these uncertainties can be difficult to
assess.

\begin{figure}
\epsscale{1.0}
\plotone{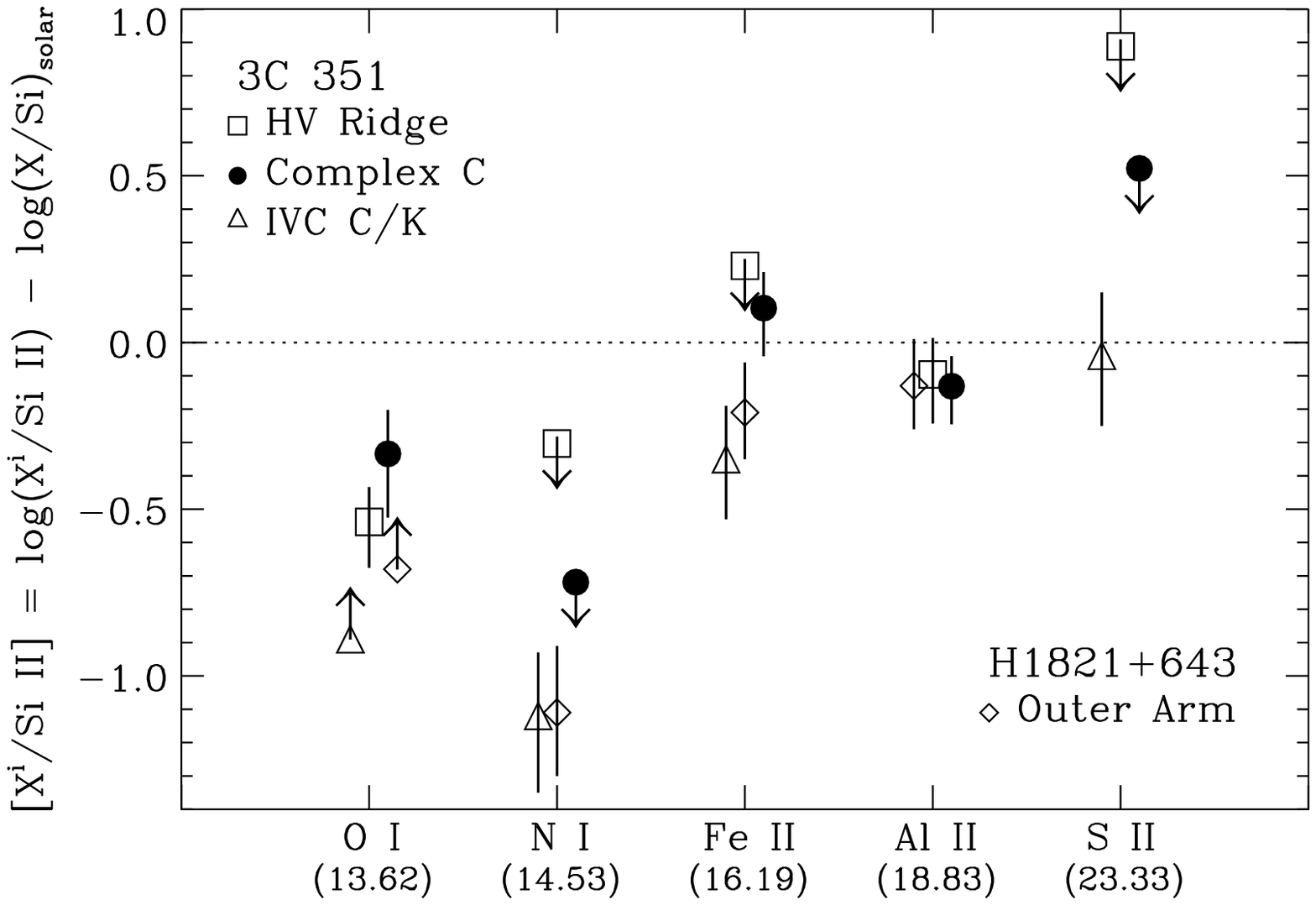}
\caption[]{Logarithmic abundances, with respect to {\it
silicon}, of \ion{O}{1}, \ion{N}{1}, \ion{Fe}{2},
\ion{Al}{2}, and \ion{S}{2} in Complex C (filled circles),
the high-velocity ridge (open squares), and the
intermediate-velocity cloud C/K (open triangles) in the
direction of 3C 351. The abundances in the outer arm
component toward H1821+643 are also shown (open diamonds).
The species are plotted in order of increasing ionization
potential (shown in parentheses in eV below each element;
the IP of \ion{Si}{2} is 16.35 eV). No ionization
corrections have been applied. The underabundances of
\ion{O}{1} and \ion{N}{1} with respect to \ion{Si}{2}
suggest that ionization corrections are
important.\label{relabun}}
\end{figure}

If we neglect the ionization correction for the Complex C absorption 
lines toward 3C 351, we obtain highly discrepant results from 
neutral-gas tracers compared to species that can persist in ionized 
gas, i.e., the different ions imply significantly different 
metallicities. For example, we obtain [\ion{O}{1}/\ion{H}{1}] $= -0.7$ 
and [\ion{N}{1}/\ion{H}{1}] $\lesssim -1.1$ vs. 
[\ion{Si}{2}/\ion{H}{1}] $= -0.4$, [\ion{Fe}{2}/\ion{H}{1}] $= -0.3$, 
and [\ion{Al}{2}/\ion{H}{1}] = $-0.5$. If correct, these abundances 
would indicate a very unusual relative abundance pattern. It is more 
likely that these discrepancies indicate that ionization corrections 
are important for the 3C 351 sight line.
These abundance discrepancies are larger than the uncertainties in the 
respective measurements, as shown in Figure~\ref{relabun}. In this 
figure we plot relative abundance measurements or limits, with respect 
to \ion{Si}{2}, for Complex C, the high-velocity ridge, and IVC C/K 
derived directly from the \ion{O}{1}, \ion{N}{1}, \ion{Fe}{2}, 
\ion{Al}{2}, and \ion{S}{2} lines, i.e., without ionization 
corrections. We plot abundances relative to \ion{Si}{2} because we do 
not have a robust $N$(\ion{H}{1}) measurement in the high-velocity 
ridge or IVC C/K, and because \ion{Si}{2} is the best-constrained 
species in Complex C. The abundances in Figure~\ref{relabun} are 
arranged according to ionization potential. Again, we see evidence of 
ionization effects in all three of these clouds: the neutral species 
\ion{O}{1} and \ion{N}{1} are significantly underabundant with respect 
to \ion{Si}{2} while 
ionized-gas tracers (\ion{Fe}{2} and \ion{Al}{2}) are much closer to 
the solar pattern.

\subsection{Complex C Proper}

\subsubsection{Collisional Ionization\label{seccollion}}

\begin{figure}
\plotone{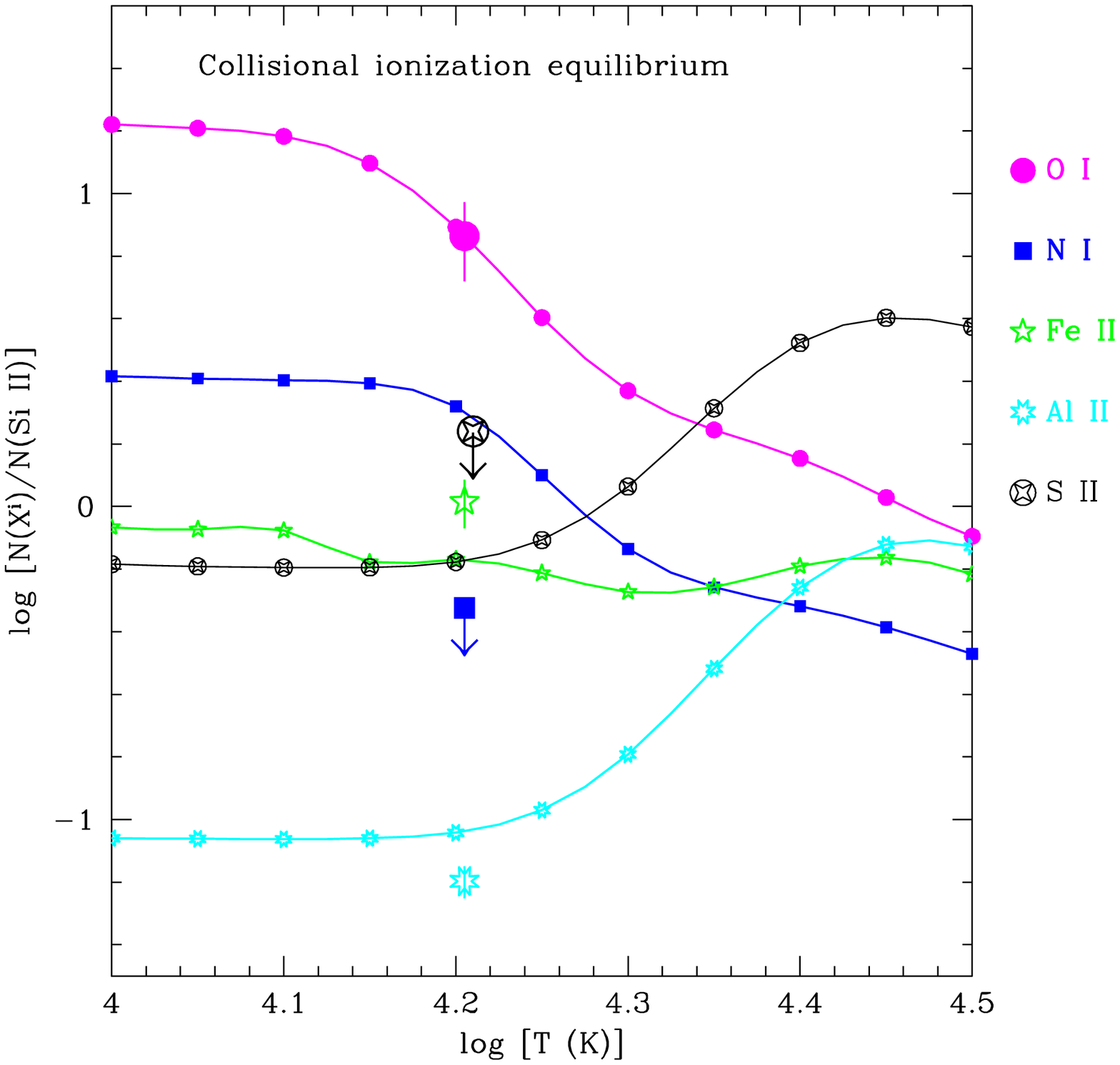}
\caption[]{Column density ratios predicted in collisional
ionization equilibrium vs. gas temperture, based on the
calculations of Sutherland \& Dopita (1993) with solar
relative abundances from Holweger (2001) or Grevesse et al.
(1996). Ratios of various species to $N$(\ion{Si}{2}) are
shown with small symbols according to the key on the right
side. Observed ratios are plotted at log $T = 4.2$ with
larger symbols and 1$\sigma$ error bars. Arrows indicate
4$\sigma$ upper limits.\label{collmod}}
\end{figure}

Can we reconcile the various abundances by applying ionization 
corrections? Since we have reasons to believe that collisional 
processes may be important (see \S1), and in order to make our 
development more clear, we begin with a simplified picture where 
collisional ionization equilibrium applies, with no additional 
photoionization from an energetic radiation field (we will add 
photoionization in the next section). Figure~\ref{collmod} shows 
relevant column density ratios, as a function of temperature, derived 
from the equilibrium collisional ionization calculations of Sutherland 
\& Dopita (1993) with solar reference abundances from Holweger (2001) 
and Grevesse et al. (1996).\footnote{Most elements in the Sutherland \& 
Dopita (1993) calculation begin in the dominant ionization stage in 
H~I regions at the lowest temperature considered. However, iron 
is an exception; the Fe~II ion fraction is significant at the 
lowest $T$. We have assumed that Fe starts from Fe~II in the 
ionization ladder because any realistic radiation field will fully 
ionize the Fe~I.} The observed ratios in Complex C are 
overplotted for comparison. We see that over the temperature range in 
which the \ion{O}{1}/\ion{Si}{2} ratio is reproduced (within 
1$\sigma$), the model does not exactly match the other observed ratios; 
departures from solar relative abundances are required. Nitrogen, for 
example, must be made underabundant with respect to silicon by {\it at 
least} $0.5 - 0.7$ dex (an underabundance moves the model curves down 
in Figure~\ref{collmod}). This may not be surprising since there is 
other evidence that nitrogen is underabundant in Complex C (Richter et 
al. 2001; Collins et al. 2003), and nucleosynthesis in low-metallicity 
gas can result in N underabundances (e.g., Henry et al. 2000). 
Similarly, Figure~\ref{collmod} indicates that aluminum must be 
underabundant if the gas is collisionally ionized and in equilibrium. 
Like nitrogen, Al could be underabundant, although much less so, due to 
nucleosynthesis effects since it is an odd-Z element, as observed in 
low-metallicity Galactic field stars (e.g., Lauroesch et al. 1996). An 
Al underabundance in the gas phase could alternatively be due to 
depletion by dust since aluminum is highly refractory (e.g., Jenkins 
1987). However, this would be inconsistent with the iron abundance: 
\ion{Fe}{2} should also be depleted in this case, but 
Figure~\ref{collmod} seems to require a slight Fe overabundance, by 
$\sim$0.2 dex, which leaves little room for dust depletion. 

\begin{figure}
\plotone{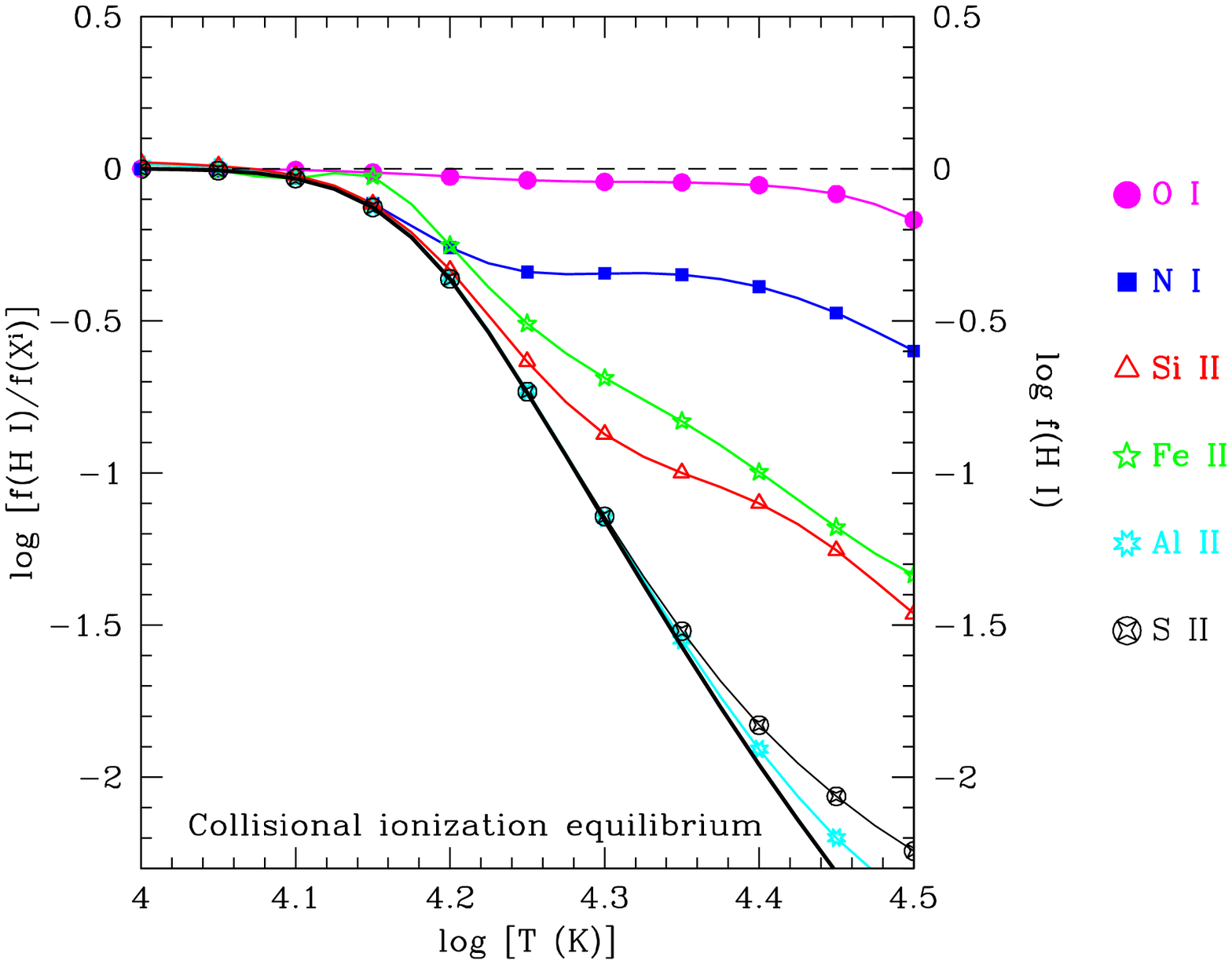}
\caption[]{Ionization corrections for collisionally ionized
gas in equilibrium from the calculations of Sutherland \&
Dopita (1993) vs. log $T$. As in Figure~\ref{collmod}, the
species corresponding to each curve is indicated by the key
at the right. The \ion{H}{1} ion fraction ($f$ =
\ion{H}{1}/H$_{\rm total}$) is shown with a heavy black
line using the scale on the right axis. The \ion{Al}{2} and
\ion{S}{2} curves are very similar because over this
temperature range, $f$(\ion{Al}{2}) $\approx f$(\ion{S}{2})
$\approx 1$, so the ionization correction is nearly equal
to $f$(\ion{H}{1}).\label{collion}}
\end{figure}

The only puzzling implication of Figure~\ref{collmod} is, in fact, the 
Fe overabundance. Nitrogen and Al underabundances would imply, at face 
value, that the gas is relatively pristine. However, in this case Fe 
should be {\it underabundant} as well since iron is thought to be 
mainly synthesized on longer timescales in Type Ia supernovae. We note 
the \ion{Fe}{2} profile is the noisiest measurement among the detected 
lines (compare Figure~\ref{navprofs}c to the other panels in 
Fig.~\ref{navprofs}), and the observed \ion{Fe}{2}/\ion{Si}{2} ratio is 
only 2$\sigma$ above the model prediction at log $T \sim$ 4.2. The 
slight Fe overabundance may simply be a result of noise. In fact, we 
shall see below that when uncertainties in the reference abundances as 
well as the column density measurements are taken into account and 
ionization corrections are applied, the final abundances agree within 
1$\sigma$, except nitrogen. However, it is clear that iron is not 
highly underabundant due to dust depletion or nucleosynthesis effects. 
We also note that Murphy et al. (2000) find a relatively high 
\ion{Fe}{2}/\ion{H}{1} ratio for Complex C in the direction of Mrk 876. 
For Figure~\ref{collmod}, we assumed solar relative abundances. If we 
were to adopt overabundances of the 
$\alpha-$elements (\ion{O}{1}, \ion{Si}{2}, and \ion{S}{2}) by 
$\sim$0.3 dex, on the grounds that such patterns are observed in 
low-metallicity stars (McWilliam 1997, and references therein), then we 
would have a substantial discrepancy between the observed and model 
\ion{Fe}{2} columns.

For the calculation of abundances using equation~\ref{metlim}, we 
present in Figure~\ref{collion} the ionization corrections obtained 
from the collisional ionization equilibrium calculations of Sutherland 
\& Dopita (1993). The \ion{O}{1}/\ion{Si}{2} ratio suggests that log $T 
\approx$ 4.20; taking the ion fractions from Figure~\ref{collion} at 
this temperature, with the column densities from Table~\ref{lineprop} 
and $N$(\ion{H}{1}) = $(4.2 \pm 1.5) \times 10^{18}$ cm$^{-2}$, we 
obtain the following abundances for Complex C proper:
\begin{equation}
[{\rm O/H}]_{\rm C} = -0.76^{+0.23}_{-0.21},
\end{equation}
\begin{equation}
[{\rm N/H}]_{\rm C} \leq -1.37,   
\end{equation}
\begin{equation}
[{\rm Si/H}]_{\rm C} = -0.73^{+0.20}_{-0.14},
\end{equation}
\begin{equation}
[{\rm Fe/H}]_{\rm C} = -0.54^{+0.22}_{-0.17},
\end{equation}
\begin{equation}
[{\rm Al/H}]_{\rm C} = -0.88^{+0.21}_{-0.15},
\end{equation}
and
\begin{equation}
[{\rm S/H}]_{\rm C} \leq -0.31.
\end{equation}
The error bars in these abundances include the uncertainties in the
column densities and the solar reference abundances but do not include
the uncertainties in the ionization correction. {\it We note that
  application of the ionization correction removes the abundance
  discrepancies discussed above. After the ionization correction at
  log T = 4.2 has been applied, all of the abundances agree within the
  1$\sigma$ uncertainties, with the notable exception of nitrogen. The
  implied metallicity is Z = $0.1 - 0.3$ Z$_{\odot}$. The nitrogen
  underabundance suggests that the gas has undergone relatively few
  cycles of metal enrichment from stars.} The magnitude of the
nitrogen discrepancy decreases if we use the lower value for
$N$(\ion{O}{1}) from $N_{\rm a}(v)$ integration (see
Table~\ref{lineprop}), and the best estimate of [O/H] decreases as
well, by $\sim$ 0.2 dex. However, since we only have an upper limit on
$N$(\ion{N}{1}), we would still be faced with a significant nitrogen
deficit. It would be valuable to obtain additional observations of 3C
351 to more securely constrain the nitrogen situation (see additional
discussion in \S~\ref{pg1259met}), and to measure the Fe abundance
with less noise.

The combination of low overall metallicity and a nitrogen
underabundance argue that Complex C is {\it not} ejecta produced in a
Galactic fountain, which would produce gas with a higher metallicity.
One might suppose that the HVC could still be part of a Galactic
foutain if during the foutain cycle the gas is substantially diluted
with low-metallicity gas or if the fountain flow started at large
Galactocentric radii, as suggested by Gibson (2002). However, both of
these hypotheses have trouble explaining the nitrogen underabundance,
because N is not underabundant in the disk ISM (e.g., Meyer, Cardelli,
\& Sofia 1997), even at larger Galactocentric radii (Afflerbach,
Churchwell, \& Werner 1997). It seems more likely that Complex C has
an extragalactic origin. It could be stripped gas from a satellite
galaxy, or it could have a more distant origin.

\subsubsection{Photoionization\label{secphot}}

Although there is strong evidence that collisional processes are 
important in HVCs (see \S1), it is worthwhile to investigate how 
additional photoionization could alter the observed column densities. 
In addition, it remains possible that the H$\alpha$ emission and 
high-ion absorption arise from a collisionally ionized phase (e.g., an 
interface on the surface of the HVC) while the low ionization lines 
originate inside the cloud where the gas is mainly photoionized. 

For this purpose, we have constructed photoionization models using 
CLOUDY (v94.0; Ferland et al. 1998). The character of the radiation 
field to which an HVC is exposed is not entirely clear. This depends on 
the location of the HVC and the fraction of the ionizing photons that 
escape from the Galactic disk (Weiner et al. 2002; Bland-Hawthorn \& 
Maloney 2002). However, the extragalactic UV background from quasars 
and active galactic nuclei  provides a floor; additional photons from 
local sources only add to this background. Consequently, we begin with 
the UV background from QSOs+AGNs only. We model the Complex C absorber 
as a plane-parallel, constant-density slab exposed to the QSO 
background at $z = 0$ from Haardt \& Madau (1996) with $J_{\nu} = 1 
\times 10^{-23}$ ergs s$^{-1}$ cm$^{-2}$ Hz$^{-1}$ sr$^{-1}$ at 1 
Rydberg (see Weymann et al. 2001 and references therein for 
observational constraints on the UV background intensity). The absorber 
thickness is adjusted to reproduce the observed $N$(\ion{H}{1}), and 
the metallicity and ionization parameter $U$ (= H ionizing photon 
density/total H number density) are varied to match the metal column 
densities. It is important to note that the \ion{H}{1} column is high 
enough so that self-shielding effects are important. These models 
should not be simply scaled for use on higher (or lower) 
$N$(\ion{H}{1}) absorption systems. 

Of course, these models are necessarily simplified compared to a real
absorption system. However, a more detailed treatment of radiation
transfer, geometry, and multiphase effects in the model presented by
Kepner et al. (1999) in the end predicts very similar column densities
to an analogous CLOUDY model.  While this is only one test case, and
more detailed modeling is highly warranted, the good agreement with
CLOUDY is encouraging.

\begin{figure}
\plotone{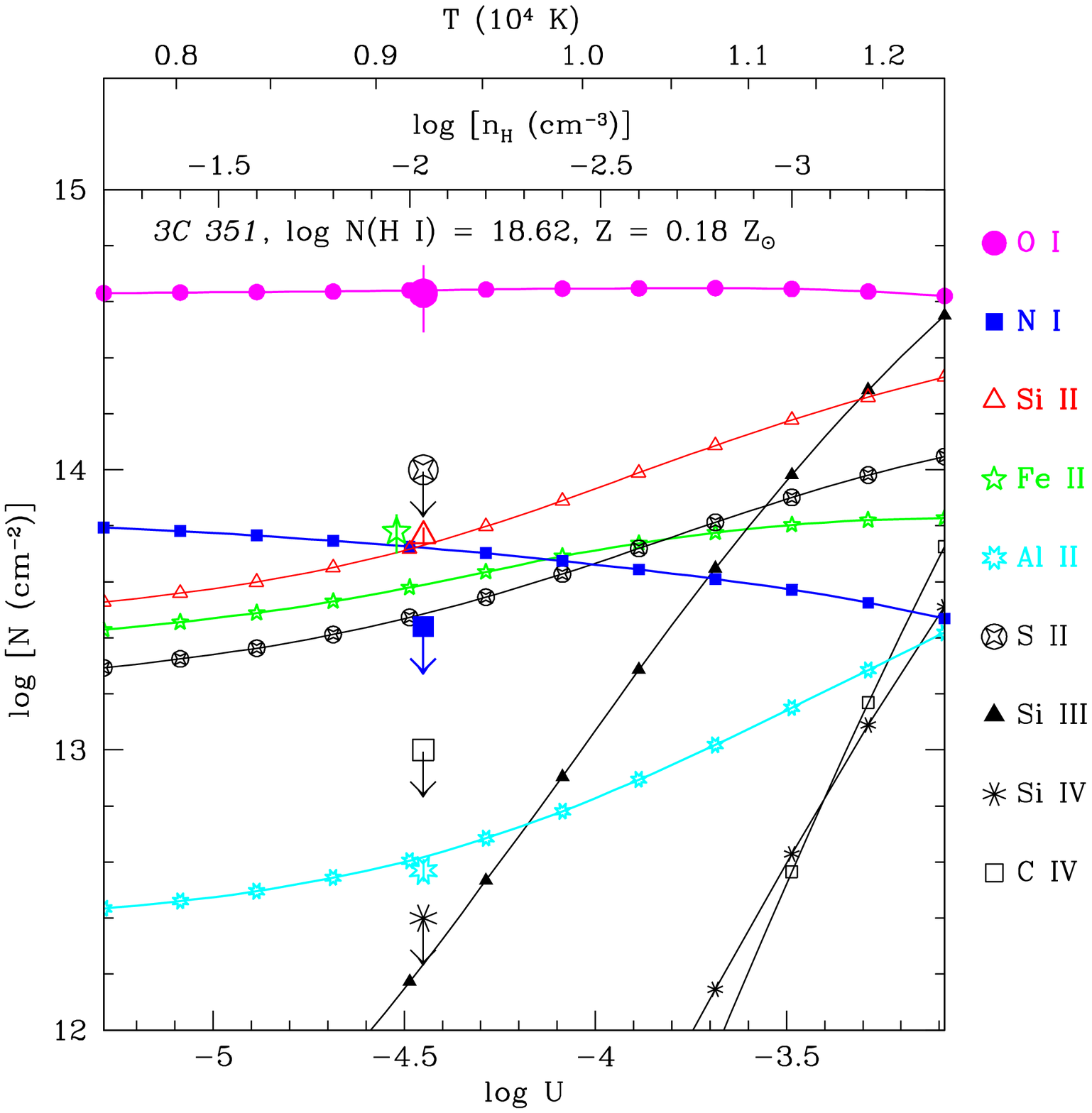}
\caption[]{Model of gas photoionized by the extragalactic
UV background at $z \sim$ 0 with log $N$(\ion{H}{1}) =
18.62, Z = 0.18Z$_{\odot}$, and relative abundances
following the solar pattern from Holweger (2002). Model
column densities are plotted with small symbols, shown in
the key at the right, as a function of the ionization
parameter $U$ (lower axis) and the particle density (middle
upper axis). The mean gas temperature is shown on the
uppermost axis; note that the temperature scale is not
linear. Observed column densities are indicated with larger
symbols with 1$\sigma$ error bars. Points with arrows are
4$\sigma$ upper limits.\label{cloudymd}}
\end{figure}

\begin{figure}
\plotone{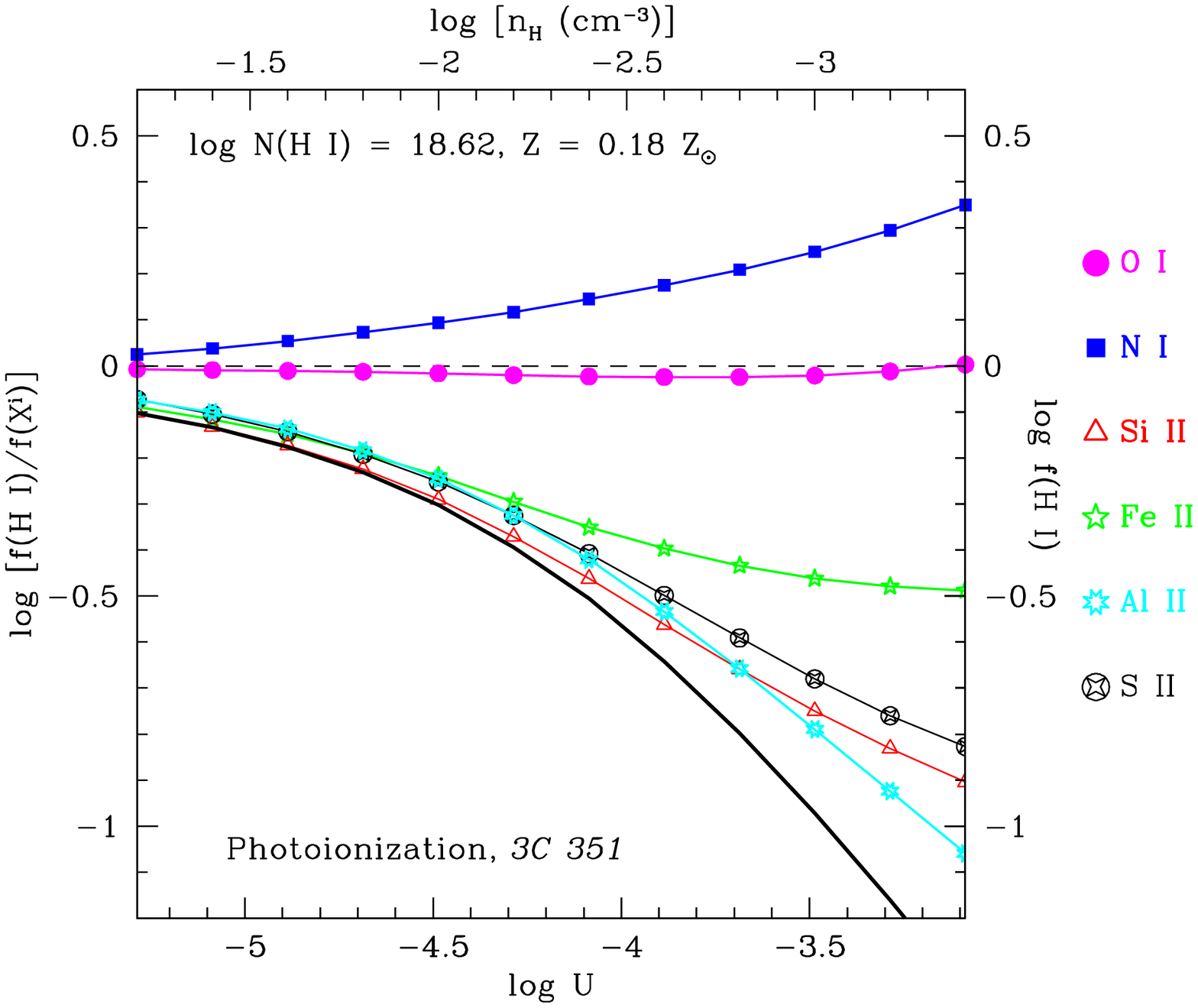}
\caption[]{Ionization corrections for the photoionization
model shown in Figure~\ref{cloudymd}. As in
Figure~\ref{collion}, the curves are identified at right,
and the thick black line indicates $f$(\ion{H}{1}) using
the right axis.\label{cloudyion}}
\end{figure}

Figure~\ref{cloudymd} shows the metal column densities predicted by
the CLOUDY model as a function of $U$, and Figure~\ref{cloudyion}
shows the corresponding ionization corrections from the same model
over the same ionization parameter range. The best fit is obtained
with log $U \approx -4.45$, as shown in Figure~\ref{cloudymd}. At this
value of $U$, the derived abundances are nearly identical to those
derived above from the collisional ionization model. The only
substantial difference between the photoionized and collisionally
ionized models is for nitrogen. In the collisionally ionized model,
all of the ionization correction factors of interest decrease as $T$
increases and the gas becomes more highly ionized (see
Figure~\ref{collion}). In the photoionized model, on the other hand,
the N ionization correction increases while the other ionization
corrections decrease as the gas becomes more highly ionized (see
Figure~\ref{cloudyion}); this is due to the relatively high
photoionization cross section of \ion{N}{1}. This reduces the nitrogen
underabundance implied by the measurements somewhat compared to the
collisional model. However, the best-fitting photoionized model still
suggests an N underabundance: we find [N/H]$_{\rm C} \leq -1.01$ vs.
[O/H]$_{\rm C} = -0.75^{+0.17}_{-0.29}.$ If we take the lower
$N$(\ion{O}{1}) from direct integration (see Table~\ref{lineprop} and
\S \ref{secmeas}), then the best fit is obtained with a somewhat
higher ionization parameter, log $U \approx - 3.95$ (log $n_{\rm H}
\approx -2.55$). As in the collisionally ionized model, this requires
a lower [O/H] by $\sim 0.2$ dex, and the nitrogen deficiency is
reduced. The iron problem is also present in the photoionized
calculation: if we make $\alpha -$elements overabundant in the model,
then we find that the observed $N$(\ion{Fe}{2}) is substantially
greater than the predicted column density. This would be unexpected in
low-metallicity gas, especially if nitrogen is underabundant.

The similarity of the photoionization and collisional ionization
results is not necessarily surprising. CLOUDY includes collisional
processes. The gas temperature in the photoionization model is
governed by the balance of photoheating and various sources of
cooling, and with the right combination of temperature and density,
the collisional processes may dominate. The different behavior of N in
the photoionized vs. collisionally ionized models is also not
surprising. Nitrogen has a relatively large photoionization cross
section, and the nitrogen + hydrogen charge exchange reaction is much
weaker than that of oxygen. Consequently, \ion{N}{1} is more readily
{\it photo}ionized than most species, and \ion{N}{1} is a sensitive
indicator of partially ionized gas (Sofia \& Jenkins 1998). We
experimented with other radiation fields in the photoionization model
with various amounts of stellar flux added to the QSO background, and
we found that the low-ion results did not change dramatically.

\subsection{High-Velocity Ridge\label{sechvr}}

The low-ion column density ratios in Complex C proper and 
the high-velocity ridge are strikingly alike (see 
Figures~\ref{navprofs}-\ref{relabun} and 
Table~\ref{lineprop}). This suggests that the physical 
conditions and abundances in Complex C and the HVR are 
quite similar. As discussed in \S2, the \ion{H}{1} column 
in the 
high-velocity ridge is not securely measured, but if we 
take the Green Bank measurement in Table~\ref{hi21} as an 
upper limit on the HVR $N$(\ion{H}{1}), we obtain 
[O/H]$_{\rm HVR}$ $\gtrsim -1.24$ (the ionization 
correction should have little effect on this [O/H] 
estimate, as shown in Figures~\ref{collion} and 
\ref{cloudyion}). This is consistent with the absolute 
metallicity derived for Complex C. However, the HV ridge 
contains substantially more highly ionized gas relative to 
the lower ionization stages. At the gas temperature and/or 
ionization parameter implied by the 
low-ion ratios in the HVR, \ion{Si}{4} and \ion{C}{4} have 
very small ion fractions and should be undetectable (see 
Table 5 in Sutherland \& Dopita 1993 and 
Figure~\ref{cloudymd}). Consequently, we conclude that the 
significant high ion absorption lines observed in the 
high-velocity ridge originate in a separate phase from the 
low ionization stages. The similar component structure of 
the low and high ions in the HVR (see Figure~\ref{stack} 
and Figure~\ref{navprofs}e-f) suggests that there is some 
association between the 
low-ionization and high-ionization phases; this could occur 
if the high ions arise in an interface between the 
low ion-bearing gas and a hotter ambient medium.

\subsection{IVC Complex C/K\label{secivc}}

The absorption-line measurements for the IVC C/K are 
substantially more uncertain than the HVC measurements due 
to line saturation and blending with adjacent components. 
Nevertheless, the 3C 351 spectrum indicates that C/K has a 
higher metallicity than Complex C proper and the HV ridge. 
The equivalent widths and column densities are 
significantly higher in the C/K component than in the 
high-velocity components (compare Tables~\ref{lineprop} and 
\ref{ivcprop}) despite the fact that the \ion{H}{1} column 
is lower in C/K than in Complex C. Adopting the \ion{O}{1} 
column from profile fitting (which provides better 
compensation for blending and saturation) and taking 
$N$(\ion{H}{1}) $< 4 \times 10^{18}$ cm$^{-2}$ (see 
\S~\ref{sec21cm}), we find that [O/H]$_{\rm C/K} > -0.5$. 
As we have shown in the previous sections, ionization 
corrections are likely to be important for the other 
species detected in C/K. We have modeled the ionization of 
the C/K gas using CLOUDY, and we find that the large 
deficit of \ion{N}{1} with respect to \ion{Si}{2} shown in 
Figure~\ref{relabun} is predominantly due to ionization. At 
the ionization parameter that provides the best fit to the 
nominal \ion{O}{1}/\ion{Si}{2} ratio (i.e., log $U = -
3.6$), the implied nitrogen underabundance is small and 
marginally significant, [N/O] $= -0.2^{+0.2}_{-0.3}$. The 
ionization-corrected abundances of the other detected 
species at this value of $U$ are listed in the last column 
of Table~\ref{ivcprop}. All abundance estimates in 
Table~\ref{ivcprop} are listed as lower limits since we 
only have an upper limit on $N$(\ion{H}{1}) 
(\S~\ref{sec21cm}).

\section{Discussion}

The measurements and modeling presented in the previous 
sections have some interesting implications. We begin our 
discussion with some comments on the structure and 
confinement of Complex C (\S 6.1). We then compare the 3C 
351 sight line to other nearby sight lines through this HVC 
(\S 6.2). We examine the case for metallicity variations in 
Complex C (\S 6.2.1), and then present evidence that the 
lower latitude region of Complex C is interacting more 
vigorously with the ambient medium (\S 6.2.2). Finally, we 
investigate the relationship between intermediate- and 
high-velocity gas observed toward H1821+643 and the HVCs in 
the 3C 351 spectrum (\S 6.2.3).

\subsection{The Structure and Confinement of Complex 
C\label{secstructure}}

Our ionization models provide constraints on the physical 
conditions and dimensions of the absorber. The size of the 
absorber along the line-of-sight is $L = N_{\rm H}/n_{\rm 
H}$. For the high (low) values of $N$(\ion{O}{1}) in 
Complex C from Table~\ref{lineprop}, the CLOUDY models that 
provide the best fits to the Complex C column densities 
(see \S~\ref{secabun}) have $n_{\rm H} = 9 \times 10^{-3} \ 
(3 \times 10^{-3})$ cm$^{-3}$, and $L = 0.30 \ (2.0)$ kpc, 
and $T \approx 9300 \ (10,700)$ K. These densities and 
temperatures suggest that the cloud may not be 
gravitationally confined. Using eqn. 3 from Schaye (2001), 
we see that a self-gravitating cloud with these $n_{\rm H}$ 
and $T$ values would have a much larger size, $L = 5.3 \ 
(10)$ kpc. The self-gravitating cloud size may change by a 
small amount depending on the absorber geometry, as noted 
by Schaye. However, even with this small scale factor, the 
uncertainties in the parameters derived from the CLOUDY 
calculations do not appear to be sufficient to reconcile 
the large difference between the expected size for a 
self-gravitating cloud and the size implied by the 
ionization models. 

The HVC could still be self-gravitating if it is predominantly
composed of dark matter. Following Schaye (2001), we have assumed that
the fraction of the mass in gas $f_{g} \approx \Omega_{b}/\Omega_{m} =
0.16$ to calculate the self-gravitating cloud size above. The cloud
could be made self-gravitating by reducing $f_{g}$ to the order of
$10^{- 2} - 10^{-4}$. This may be unlikely, but it is not
inconceivable. As an HVC plunges into a galaxy halo, ram pressure can
separate the gas from its original dark matter halo (e.g., Quilis \&
Moore 2001).  We may be viewing the small amount of residual gas in a
dark matter halo that has already had most of its baryons stripped
away. However, Quilis \& Moore find that this requires the ambient
medium to have a relatively high density, $\gtrsim 10^{-4}$ cm$^{-3}$,
which they deem ``unrealistic for a Galactic halo component''.
However, their analysis mainly considers HVCs at distances of
$\gtrsim$ 100 kpc. If Complex C is only $\sim$10 kpc away, the ambient
halo density could be much higher, making this hypothesis more
plausible.  We note that the requirement of a small value for $f_{g}$
to make the cloud self-gravitating does not strictly require {\it
  dark} matter.  However, searches for other components in HVCs such
as stars (e.g., Willman et al.  2002) or molecular hydrogen (e.g.,
Richter et al.  2002) have not produced detections despite good
sensitivity.

On the other hand, we may be observing the other side of this process,
i.e., our gas may have been recently stripped out of a dark matter
halo and is now largely devoid of dark matter.  In this case, there is
no particular need to confine the gas; the stripped gas could then be
an ephemeral cloud that will rapidly evaporate in the hot halo. 

The purely collisionally ionized model (\S 
\ref{seccollion}) does not provide a direct constraint on 
the number density and hence the size of the absorber, but 
we argue that the situation is similar. The collisional 
model does provide an estimate of the \ion{H}{1} ion 
fraction $f_{\rm H~I}$. Substituting 
\begin{equation}
n_{\rm H} = \frac{N({\rm H~I})}{f_{\rm H~I} L} 
\label{sgdensity}
\end{equation}
into Shaye's eqn. 3, we obtain
\begin{equation}
L = 8.1 f_{\rm H~I} \left(\frac{T}{10^{4} \ {\rm K}}\right) 
\left( \frac{N({\rm H~I})}{1 \times 10^{20} \ {\rm cm}^{-
2}} \right)^{-1} \left( \frac{f_{g}}{0.16}\right) \ {\rm 
kpc}. \label{sgsize}
\end{equation}
The best-fit from the collisional ionization model with the 
high $N$(\ion{O}{1}) has $f_{\rm H~I}$ = 0.43 and $T = 
10^{4.2}$ K, so with $N$(\ion{H}{1}) = $4.2 \times 10^{18}$ 
we derive $L = 130$ kpc and $n_{\rm H} = 2.5 \times 10^{-
5}$ cm$^{-3}$ from equations~\ref{sgdensity} and 
\ref{sgsize} for a gravitationally confined cloud. The 
angular extent of Complex C (roughly $20^{\circ} \times 
90^{\circ}$) implies that the transverse size of the entire 
complex is $\sim 3.5 \times 20$ kpc if the cloud is 10 kpc 
away. Therefore $L = 130$ kpc requires an unlikely 
geometry: the cloud must be vastly larger along the 
line-of-sight than in the transverse direction. We reach 
the same conclusion with the parameters resulting from the 
lower \ion{O}{1} column. This discrepancy could be 
reconciled, once again, by invoking a relatively small 
fraction of the mass in gas, $f_{g}$. The implied geometry 
would also be more plausible if Complex C was considerably 
farther than 10 kpc, but there are arguments against 
placing this HVC much farther away (Wakker et al. 1999a; 
Blitz et al. 1999). We conclude that the collisional 
ionization equilibrium model faces the same problems as the 
CLOUDY model.

Of course, there are alternatives to gravitational confinement, and
there is no requirement for HVCs to contain dark matter at all.
Magnetic confinement may be important (Konz, Br\"{u}ns, \& Birk 2002).
Pressure confinement by a hotter external medium is also a reasonable
alternative.  The widespread detection of high-velocity \ion{O}{6}
absorption (Sembach et al. 2002) suggests that the Milky Way has a
large, hot corona. Such a corona could provide the pressure
confinement that we require. The densities and temperatures from the
best CLOUDY models imply that the gas pressure in the Complex C
component is $p/k \approx 60 - 170$ cm$^{-3}$ K. If the external
confining medium has $T_{\rm ext} \gtrsim 10^{6}$ K, then its density
$n_{\rm ext} \lesssim 6 \times 10^{-5} - 2 \times 10^{-4}$ cm$^{- 3}$
in order to pressure confine the HVC. This density upper limit is in
agreement with density constraints from other arguments (e.g., Wang
1992; Moore \& Davis 1994; Weiner \& Williams 1996; Murali 2000;
Br\"{u}ns, Kerp, \& Pagels 2001). We note that these pressures are
also consistent with constraints derived by Sternberg, McKee, \&
Wolfire (2002) on the pressure in the intragroup medium in the Local
Group (based on \ion{H}{1} structure properties in nearby dwarf
galaxies).

We can apply the same analysis to the high-velocity ridge 
component, but we only obtain limits since we only have an 
upper limit on $N$(\ion{H}{1}). For this high-velocity 
component, the best-fitting CLOUDY model implies an 
absorber thickness $L \leq$ 2.6 kpc (decreasing 
$N$(\ion{H}{1}) also decreases $L$). With $n_{\rm H} 
\approx$ 2.5$\times 10^{-3}$ cm$^{-3}$ and $T \approx 
11,300$ K, the estimated size for a self-gravitating cloud 
(assuming $f_{g}$ = 0.16 again) is $L =$ 11 kpc. In this 
case, decreasing $f_{g}$ by a factor of $2-3$ reconciles 
the self-gravitating cloud size with the size implied by 
the ionization model. Gravitational confinement appears to 
be viable for the HVR, as long as the actual 
$N$(\ion{H}{1}) is not too much lower than our upper limit 
(see \S~\ref{sec21cm}). 

Summarizing this section, our constraints on the physical conditions
and dimensions of the Complex C gas toward 3C 351 indicate that it is
unlikely that this absorption arises in a gravitationally confined
cloud. It is much more probable that the gas is pressure confined by
an external medium or is an ephemeral entity that will soon dissapate.
The HVR could be gravitationally confined, but if the HVR and Complex
C proper are indeed closely related, then the physical processes
affecting Complex C are likely to affect the HVR as well, and the
situation with the HVR may be similar to that of Complex C proper.

\subsection{Sight-Line Comparisons\label{seccompare}}

\subsubsection{Mrk 279, Mrk 817, and 
PG1259+593\label{pg1259met}}

\begin{figure}
\plotone{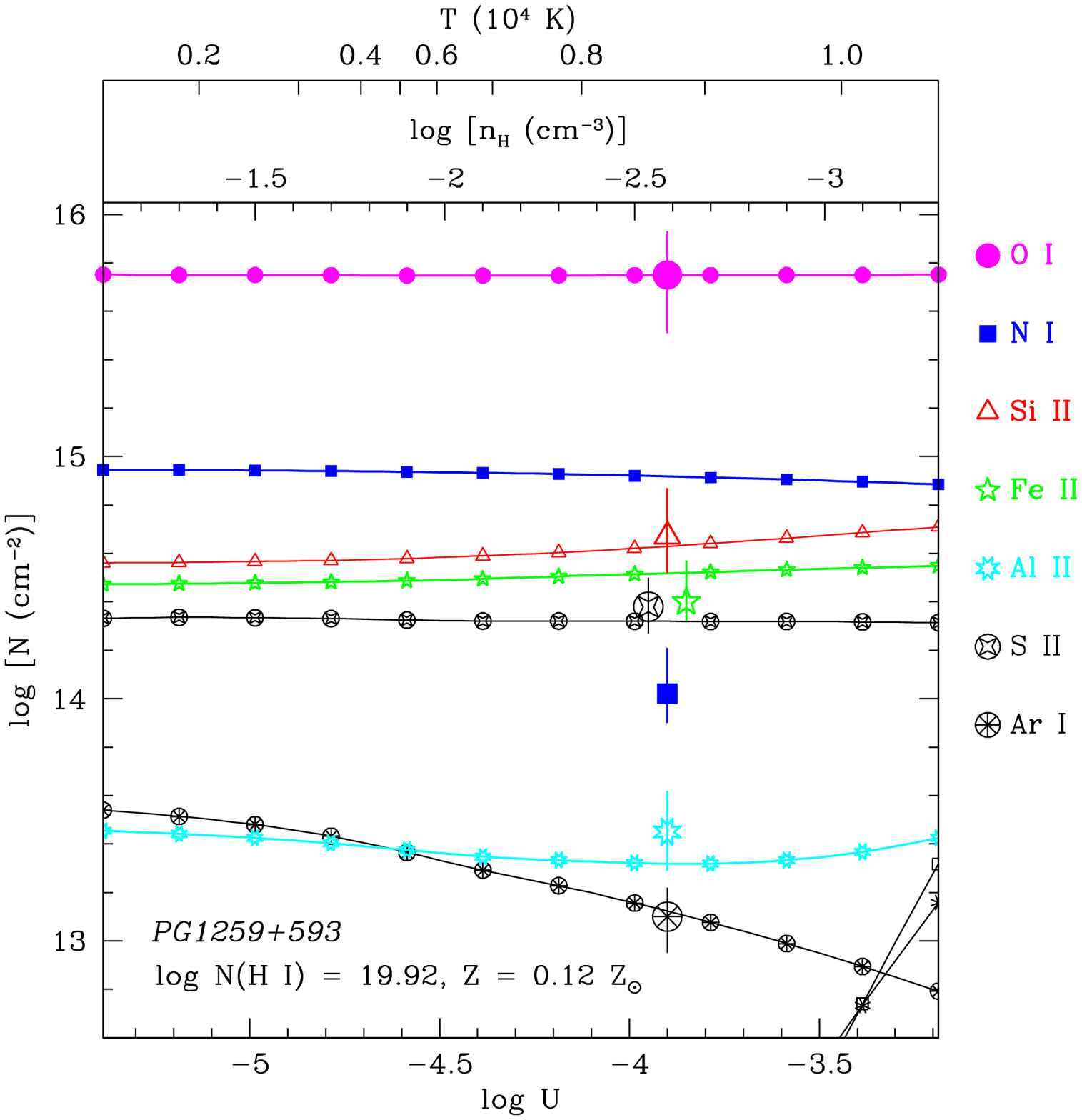}
\caption[]{Model of gas photoionized by the extragalactic
UV background, as in Figure~\ref{cloudymd}, but with
parameters appropriate for the sight line to PG1259+593:
log $N$(\ion{H}{1}) = 19.92, Z = 0.12Z$_{\odot}$, and solar
relative abundances. The large points show the observed
Complex C column densities in the direction of PG1259+593
from Collins et al. (2003). At log $U = -3.9$, all of the
observed columns are within 1$\sigma$ of the model column
densities except \ion{N}{1}.\label{cloudy1259}}
\end{figure}

We can gain additional insights by comparing the 3C 351 
results to other sight lines through the Complex C 
high-velocity cloud. Richter et al. (2001) and Collins et 
al. (2002) have published accurate column densities, 
including a variety of species, for several extragalactic 
sight lines through Complex C. These papers have emphasized 
abundance results, but the observations have implications 
regarding the physical conditions/structure of the HVC as 
well. From these papers, the spectra of Mrk 279, Mrk 817, 
and PG1259+593 have the highest S/N and provide the best 
constraints, so we shall concentrate on these sight lines.

An important difference between 3C 351 and 
Mrk279/Mrk817/PG1259+593 is that the latter sight lines 
have substantially higher \ion{H}{1} column densities. The 
increased self-shielding resulting from the higher 
$N$(\ion{H}{1}) has advantages and disadvantages: on the 
one hand, the ionization corrections are smaller, but the 
ion ratios also provide less leverage on the density and 
physical conditions of the gas. Figure~\ref{cloudy1259} 
shows an example. This figure shows a CLOUDY calculation, 
analogous to the model presented in \S \ref{secphot}, of 
metal columns vs. log $U$ and log $n_{\rm H}$ for the sight 
line to PG1259+593. The only difference from the 3C 351 
model is that the \ion{H}{1} column is higher (in accord 
with 21cm observations), and the overall metallicity has 
been reduced by a small amount to provide the best fit to 
the observed column densities. Because of the higher 
$N$(\ion{H}{1}) and greater self-shielding, most of the 
curves in Figure~\ref{cloudy1259} are relatively flat; 
$f$(X$^{i}$) = 1 over a large range of $U$ for most species 
shown. Consequently, the derived abundances are insensitive 
to $U$, but also most of the measurements allow a large 
range of $n_{\rm H}$, $L$, and $T$. However, \ion{Ar}{1} is 
a notable exception. As discussed by Sofia \& Jenkins 
(1998), \ion{Ar}{1} is susceptible to photoionization, and 
this can be useful for investigation of gas ionization. 
This can be seen in Figure~\ref{cloudy1259}. While most of 
the curves change by 0.1 dex or less, $N$(\ion{Ar}{1}) 
decreases by $\sim 0.8$ dex over the range of $U$ shown in 
the figure. Unfortunately, the argon lines are relatively 
weak, and of the eight sight lines presented by Collins et 
al. (2002), \ion{Ar}{1} is only detected toward PG1259+593.

Taking advantage of the small ionization corrections, we 
confirm the main abundance results reported by Collins et 
al. (2002), with some additional important comments:
\begin{enumerate}
\item The Complex C oxygen abundances\footnote{Our oxygen 
abundances differ from those reported in Collins et al. 
(2002) because we adopt the revised solar oxygen abundance 
reported by Holweger (2001) while Collins et al. adhere to 
the previous (O/H)$_{\odot}$ from Grevesse \& Sauval 
(1998). We also include the uncertainties in the solar 
reference abundances reported by Holweger in the overall 
error bars.} that we derive for these three directions show 
some dispersion:
\begin{equation}
[{\rm O/H}]_{\rm Mrk279} = -0.59^{+0.36}_{-0.26}, 
\label{mrk279abun}
\end{equation}
\begin{equation}
[{\rm O/H}]_{\rm Mrk817} = -0.50^{+0.25}_{-0.18}, 
\label{mrk817abun}
\end{equation}
and
\begin{equation}
[{\rm O/H}]_{\rm PG1259+593} = -0.91^{+0.20}_{-
0.25}.\label{pg1259abun}
\end{equation}
However, including the 3C 351 oxygen abundance, we find that the
weighted mean metallicity for Complex C is $<$[O/H]$>$ = $-0.68$ with
a reduced $\chi ^{2}_{\nu} = 0.85$ ($\nu = 3$). Therefore the current
measurements are consistent with a constant metallicity throughout
Complex C with $Z \approx 0.2 Z_{\odot}$. Furthermore, the error bars
in \ref{mrk279abun} - \ref{pg1259abun} do not fully reflect the
$N$(\ion{H}{1}) uncertainties. For example, the \ion{H}{1} 21 cm
profile toward Mrk279 is very complex, and it is difficult to separate
the Complex C \ion{H}{1} emission from the lower-velocity emission.
Also, the Mrk817 $N$(\ion{H}{1}) was derived from the larger-beam LDS
data, which introduces a systematic error of $0.3 - 0.5$ dex.  With
these additional uncertainties, we conclude that there is currently no
compelling evidence of oxygen abundance variations in Complex C. We
cannot rule out the possibility that the oxygen abundances are spatially
variable in Complex C, but better measurements are required to support
this claim.
\item Nitrogen is underabundant in Complex C:
\begin{equation}
[{\rm N/H}]_{\rm Mrk279} < -1.1,
\end{equation}
\begin{equation}
[{\rm N/H}]_{\rm Mrk817} < -1.2,
\end{equation}
\begin{equation}
[{\rm N/H}]_{\rm Mrk876} = -1.20^{+0.17}_{-0.15},
\end{equation}
and
\begin{equation}
[{\rm N/H}]_{\rm PG1259+593} = -1.83^{+0.22}_{-0.16},
\end{equation}
where the upper limits are at the $4\sigma$ level. Again, this
indicates that intermediate-mass stars have not contributed
significantly to the gas enrichment. The nitrogen detections toward
Mrk876 and PG1259+593 provide stronger evidence of abundance
variations in Complex C than the oxygen measurements: these two sight
lines yield a weighted mean of $<$[N/H]$>$ = $-1.40$ with a reduced
$\chi ^{2}_{\nu}$ = 5.60. Some caution is warranted, though, because
$N$(\ion{N}{1}) toward Mrk876 is entirely based on a weak and blended
line (see Figure 8 in Collins et al. 2002), and the measurement may be
more uncertain than the formal error bars suggest.\footnote{The Mrk876
  N~I column has also been estimated from the N~I
  $\lambda$1134.17 line observed with {\it FUSE} (Murphy et al. 2000).
  However, the feature at the expected velocity of N~I
  $\lambda$1134.17 in Complex C is strongly blended with low-velocity
  Fe~II $\lambda$1133.67, and in fact the line is predominantly
  due to Fe~II. Murphy et al.  (2000) have subtracted this
  Fe~II line based on other Fe~II transitions in the {\it
    FUSE} bandpass, but the resulting $N$(N~I) estimate is
  considerably uncertain.}
\item The current error bars are too large to provide 
significant evidence of $\alpha-$element overabundances. 
The model shown in Figure~\ref{cloudy1259} provides a 
satisfactory match of the observed column densities with 
solar relative abundances, for example. The fact that the 
\ion{S}{2} and \ion{Fe}{2} column densities are comparable 
suggests that the $\alpha-$elements may be overabundant, 
but within current uncertainties, both [$\alpha$/Fe] = 0 
and [$\alpha$/Fe] = +0.3 provide acceptable fits to the 
observations.
\item However, the $N$(\ion{Fe}{2})/$N$(\ion{S}{2}) ratios 
do indicate that there is little depletion by dust in 
Complex C since dust strongly reduces the gas-abundance of 
Fe but has little effect on S in a variety of environments 
(e.g., Jenkins 1987; Sembach \& Savage 1996).
\end{enumerate}

As noted above, most metals detected along these higher 
$N$(\ion{H}{1}) sight lines allow a wide range of densities 
and sizes. However, \ion{Ar}{1} is useful. Toward 
PG1259+593, the best fit to the metals in 
Figure~\ref{cloudy1259}, including \ion{Ar}{1} but 
excluding \ion{N}{1}, has $n_{\rm H} = 2.5 \times 10^{-3}$, 
$T = 8800$ K, and $L = N_{\rm H}/n_{\rm H} = 13.9$ kpc. 
This implied size is vastly larger than the size derived 
from the 3C 351 data. Furthermore, in the direction of 
PG1259+593, gravitational confinement appears to be quite 
viable: this only requires $f_{g} = 0.32$, i.e., a factor 
of two larger than Schaye's fiducial value of 0.16. The 
upper limits on Complex C $N$(\ion{Ar}{1}) for the Mrk 279 
and Mrk 817 sight lines do provide the following 
constraints. For Mrk 279, $n_{\rm H} \lesssim 5 \times 
10^{-3}$ cm$^{-3}$ and $L \gtrsim$ 2.5 kpc. For Mrk 817, 
$n_{\rm H} \lesssim 1 \times 10^{-3}$ cm$^{-3}$ and $L 
\gtrsim$ 14 kpc.

\subsubsection{3C 351: at the Leading Edge of Complex C}

\begin{figure}
\plotone{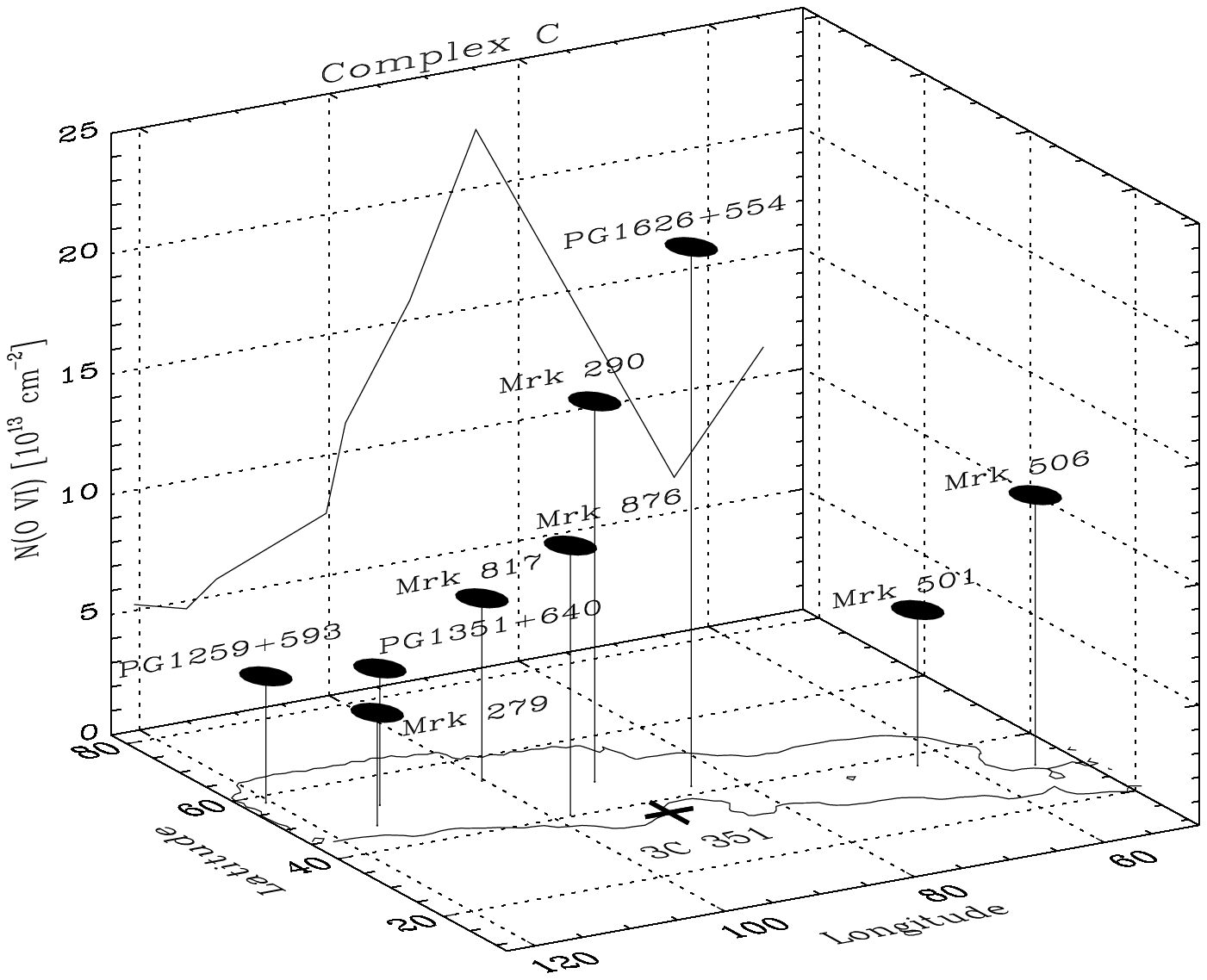}
\caption[]{Map of the \ion{O}{6} column density in Complex
C proper. The xy-plane shows the $N$(\ion{H}{1} = $2 \times
10^{18}$ cm$^{-2}$ contour for Complex C from Hulsbosch \&
Wakker (1988), plotted vs. Galactic longitude and latitude.
The z-axis shows $N$(\ion{O}{6}), in units of $10^{13}$
cm$^{-2}$, measured towards nine QSOs/AGNs, from Table 8 in
Sembach et al. (2002). The column densities are shown with
filled circles, with a line marking the sight line location
within Complex C in the contour plot in the xy-plane.
$N$(\ion{O}{6} is also projected onto the xz-plane. The
location of the 3C 351 sight line is marked with a {\bf +};
\ion{O}{6} cannot be measured in the 3C 351 spectrum
because of an extragalactic Lyman limit absorber at $z_{\rm
abs} = 0.221$.\label{ovimap}}
\end{figure}

\begin{figure}
\plotone{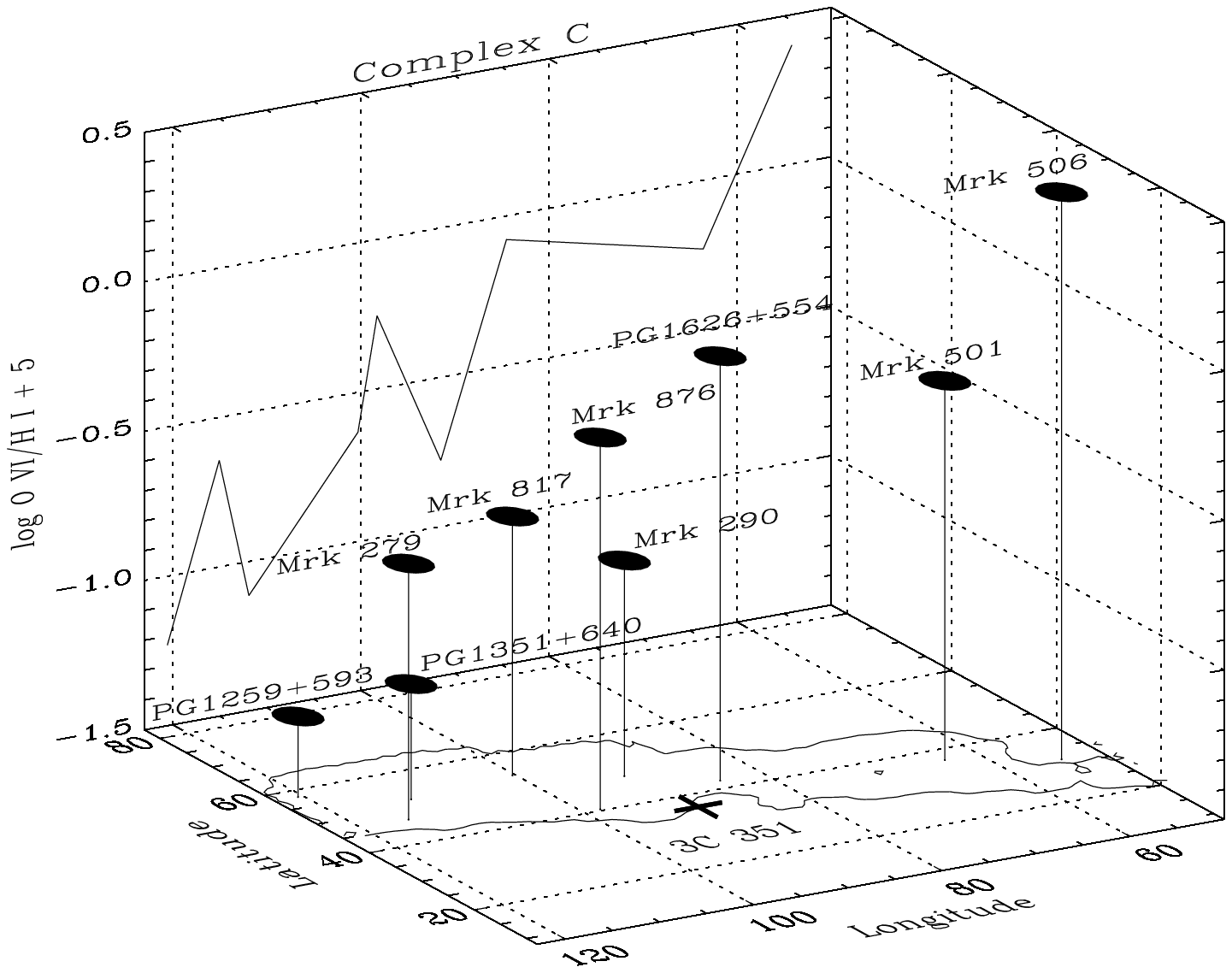}
\caption[]{Map of the \ion{O}{6}/\ion{H}{1} ratio in
Complex C proper, as in Figure~\ref{ovimap}. The z-axis
shows the logarithm of $N$(\ion{O}{6})/$N$(\ion{H}{1})
$\times 10^{5}$. The \ion{O}{6}/\ion{H}{1} ratios are also
projected onto the xz-plane, and the location of the 3C 351
sight line is marked with a {\bf +}.\label{o6h1map}}
\end{figure}

Why do the sight lines to 3C 351 and PG1259+593 have such starkly
different implications regarding the size and confinement of the
absorber? We propose a simple answer: Complex C is probably
interacting with the thick disk/lower halo of the Milky Way much more
vigorously in the direction of 3C 351 than in the direction of
PG1259+593. This would occur naturally if the 3C 351 pencil beam is
near the leading edge of the HVC, and this would explain several
observations. First, the 21cm observations show a steep gradient in
$N$(\ion{H}{1}) in the vicinity of 3C 351 (see
Figures~\ref{hw_map21cm}-\ref{lds_compc}). This likely reflects the
transition from the mostly neutral inner region to the fully ionized
periphery of the HVC. Such a transition is expected at the leading
edge of a cloud moving through an ambient halo (e.g., Quilis \& Moore
2001). Second, ram pressure could separate the baryons from the
underlying dark matter (Quilis \& Moore 2001) and thereby alleviate
the confinement problem discussed above.  In this case there is no
compelling requirement to confine the gas; this edge of the cloud
would be rapidly evaporating as it plunges toward the disk. Third,
observations of nine sight lines through Complex C reported by Sembach
et al. (2002) show that the \ion{O}{6} column densities at the
velocities of Complex C are highest near 3C 351 (several of these
nearby sight lines are marked in
Figures~\ref{hw_map21cm}-\ref{lds_hvr}). These \ion{O}{6} observations
are summarized in Figure~\ref{ovimap}, which shows $N$(\ion{O}{6})
measured toward the nine extragalactic sources marked on a map of
Complex C. Note that these are \ion{O}{6} columns in Complex C proper
{\it only}; recall also that \ion{O}{6} cannot be measured in the 3C
351 spectrum due to a Lyman limit absorber at $z_{\rm abs} = 0.221$
that severely attenuates the FUSE 3C 351 spectrum below $\sim$1117 \AA
. As shown in Figure~\ref{o6h1map}, an even more pronounced trend is
evident in the $N$(\ion{O}{6})/$N$(\ion{H}{1}) ratio in Complex C.
This ratio increases steadily with decreasing longitude and latitude,
and the lowest-longitude Complex C sight line (Mrk 501; see Table 8 in
Sembach et al. 2002) has an $N$(\ion{O}{6})/$N$(\ion{H}{1}) ratio
which is $\sim$43 times larger than the ratio measured toward the
highest-longitude sight line (PG1259+593). This trend also suggests
that the lower-longitude section of Complex C is more strongly
interacting with the ambient medium; as the gas is ionized to a
greater degree, $N$(\ion{H}{1}) will decrease while $N$(\ion{O}{6})
increases. The lower-longitude region is also at lower latitude, so
this result is not surprising: as the HVC approaches the plane of the
Milky Way, it is becoming more fully ionized and is probably ablating
and dissipating, and the trailing (higher-longitude) portion of the
cloud has not yet entered the higher density region of the ambient
medium where the interactions are more vigorous, or at least has not
suffered the effects of ablation for the same duration as the lower
latitude/longitude regions.

\subsubsection{H1821+643\label{secouter}}

An alternative to the interpretation presented above is 
that gas flowing up out of the disk (e.g., Galactic 
fountain/chimney gas) is colliding with Complex C at lower 
longitudes and latitudes, and the resultant shock-heating 
is ionizing and evaporating the cloud. Gibson et al. (2001) 
and Collins et al. (2002) have suggested that such 
interactions might be required to explain the metallicity 
variations observed in the HVC (but see point 1 in \S 
\ref{pg1259met}). In this regard, it is interesting to 
compare the 3C 351 to the H1821+643 sight line, which is 
near 3C 351 but closer to the plane (see 
Figures~\ref{hw_map21cm}-\ref{lds_hvr}). The H1821+643 
line-of-sight passes just outside the lower-latitude 21cm 
boundary of Complex C (see also Figure 1 in Wakker \& van 
Woerden 1997) but through the region of high-velocity 
emission associated with the Outer Arm. A good spectrum of 
H1821+643 could reveal gas flowing from the plane toward 
Complex C. 

\begin{figure}
\epsscale{0.8}
\plotone{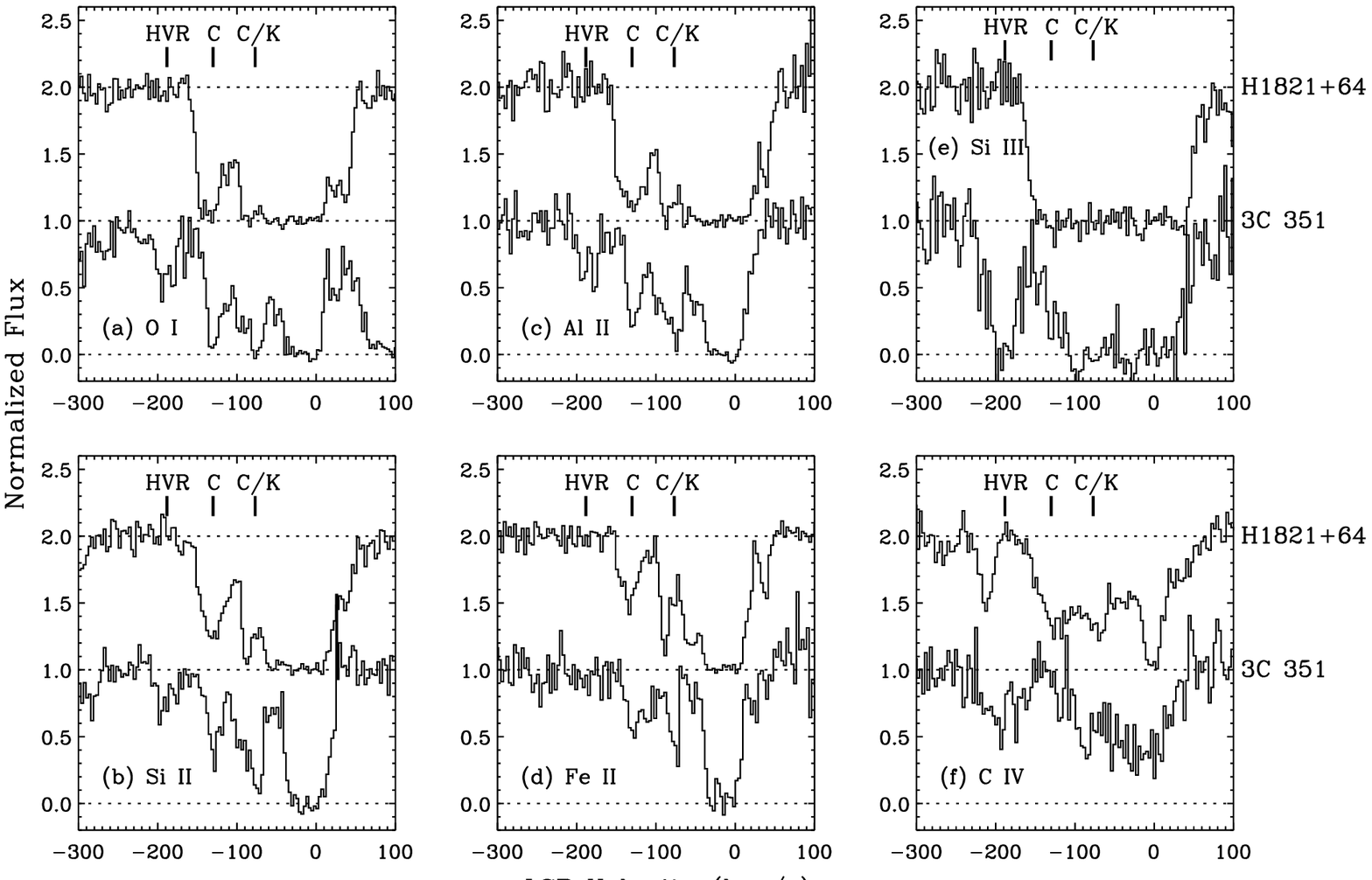}
\vspace*{1cm}
\caption[]{Comparison of normalized absorption profiles
observed in the directions of 3C 351 (lower profile in each
panel) and H1821+643 (upper profile in each panel). The
plots compare the (a) \ion{O}{1} $\lambda$1302.17, (b)
\ion{Si}{2} $\lambda$1304.37, (c) \ion{Al}{2}
$\lambda$1670.79, (d) \ion{Fe}{2} $\lambda$1608.45, (e)
\ion{Si}{3} $\lambda$1206.50, and (f) \ion{C}{4}
$\lambda$1548.20 transitions. The H1821+643 spectra have
been offset by +1 for clarity. These sight lines probe the
low-latitude edge of Complex C (see
Figures~\ref{hw_map21cm} - \ref{lds_compc} as well as
Figure 1 in Wakker \& van Woerden 1997). For reference, the
velocities of the high-velocity ridge (HVR), Complex C, and
IVC C/K are marked at the top of each
panel.\label{h1821compare}}
\end{figure}

Savage, Sembach, \& Lu (1995) have discussed Galactic 
absorption features in the spectrum of H1821+643 based on 
an intermediate-resolution GHRS spectrum. Subsequently, 
Tripp, Savage, \& Jenkins (2000) obtained a 
higher resolution STIS spectrum of H1821+643 with the same 
mode (E140M) used to observe 3C 351, so a detailed 
comparison of these sight lines can be made without 
confusion due to differing spectral resolution. 
Figure~\ref{h1821compare} compares selected absorption 
profiles observed toward H1821+643 (upper profile in each 
panel) to the same line from the spectrum of 3C 351. For 
reference, the velocities of the HVR, C, and C/K components 
are marked at the top of each panel. Some of the profiles 
appear to be remarkably similar, e.g., the \ion{Fe}{2} 
lines in Figure~\ref{h1821compare}d. We defer a full 
analysis of the H1821+643 data to a later paper, but a few 
comments on this figure are worthwhile.

{\it The v = $-$80 km s$^{-1}$ component: Complex C/K}.  Both sight
lines show components near the velocity of IVC C/K with similar
equivalent widths. Toward H1821+643, Savage et al. (1995) associate
this absorption with the Perseus spiral arm, which has a similar
velocity at $b \approx 0$ and appears to extend $\sim 1$ kpc above the
plane based on 21cm and H$\alpha$ emission (Kepner 1970; Reynolds
1986). We have derived [O/H] $> -0.5$ for the C/K component toward 3C
351, implying a higher metallicity than the adjacent Complex C and HVR
components. The higher C/K metallicity suggests that this gas may
indeed have originated in the disk and is being driven into the halo.
If the Perseus arm is $\sim 3$ kpc away and the 3C 351 C/K component
is associated with Perseus, then this gas is $\sim 2$ kpc above the
plane.  This is consistent with the $z-$heights that gas is expected
to attain in some Galactic fountain models (e.g., Houck \& Bregman
1990). Very recently, Otte, Dixon, \& Sankrit (2003) have detected
\ion{O}{6} emission at $v_{\rm LSR} = -50\pm 30$ \kms\ from $l =
95.4^{\circ}$, $b = 36.1^{\circ}$, a direction very close to the 3C
351 and H1821+643 sight lines. They also show bright H$\alpha$
filaments extending up from the plane in this direction. They
attribute both the \ion{O}{6} and H$\alpha$ emission to an outflow
from the Perseus arm. Regardless of the nature of the gas, it is quite
likely that the absorption lines at $v \approx -80$ \kms\ toward 3C
351 and H1821+643 are related to the emission reported by Otte et al.
(2003).

{\it The v = $-$130 km s$^{-1}$ component: Complex C vs. 
the Outer Arm.}
Strong absorption lines are present at the velocity of 
Complex C in the spectra of both H1821+643 and 3C 351. 
However, unlike IVC C/K, the line equivalent widths in the 
H1821+643 spectrum are substantially larger than the same 
lines toward 3C 351 at $v = -130$ km s$^{-1}$ (see 
Figure~\ref{h1821compare}). The H1821+643 and 3C 351 21 cm 
\ion{H}{1} column densities at this velocity are similar: 
Wakker et al. (2001) report $N$(\ion{H}{1}) = $3.3 \times 
10^{18}$ toward H1821+643, compared to $N$(\ion{H}{1}) = 
$4.2 \times 10^{18}$ toward 3C 351 (both measurements are 
based on Effelsberg observations). However, there is a gap 
in the 21 cm emission between the two sight lines (see 
Figures~\ref{hw_map21cm} and \ref{lds_compc}), and the 
relationship (if any) between the 3C 351 and H1821+643 HVCs 
at this velocity is unclear. We list in 
Table~\ref{outerarmprop} the equivalent widths and column 
densities of absorption lines at $v = -130$ km s$^{-1}$ in 
the H1821+643 spectrum, measured as described in 
\S~\ref{secmeas}. Several pixels in the core of the 
\ion{O}{1} line toward H1821+643 approach zero flux. This 
line may be significantly saturated, so we consider the 
\ion{O}{1} measurements highly uncertain. The most 
conservative constraint on $N$(\ion{O}{1}) is a lower limit 
from direct integration, but we also provide an estimate 
from profile fitting in Table~\ref{outerarmprop}, which can 
correct somewhat for saturation but with large 
uncertainties. We also summarize the implied abundances in 
Table~\ref{outerarmprop} as before, without and with 
ionization corrections applied (using the ionization model 
presented below).

\begin{deluxetable}{rcccllll}
\tabletypesize{\footnotesize}
\tablewidth{0pc}
\tablecaption{High-Velocity Absorption Lines toward 
H1821+643: The Outer 
Arm\tablenotemark{a}\label{outerarmprop}}
\tablehead{\ \ Species \ \ & $\lambda _{0}$ & $<v>$ & 
$W_{\lambda}$ & \ \ log $N_{\rm a}$ & \ \ log $N_{\rm pf}$ 
& \ \ [X$^{i}$/H~I]\tablenotemark{b} & \ \ 
[X/H]\tablenotemark{c} \\
 \ & (\AA ) & (km s$^{-1}$) & (m\AA ) & \ & \ }
\startdata
O~I\dotfill & 1302.17 & $-131\pm 1$ & 183.6$\pm$5.4 
& $>14.70$\tablenotemark{d} & 15.18$^{+0.26}_{-
0.24}$\tablenotemark{e} & $\geq -0.55$ & $\geq -0.57$ \\
N~I\dotfill & 1199.55 & $-136\pm 5$ & 40.0$\pm$9.3 & 
13.47$\pm 0.11$ & $13.40 \pm 0.09$ & $-0.97 \pm 0.17$ & $-
0.73 \pm 0.17$ \\
Si~II\dotfill & 1304.37 & $-128\pm 1$ & 134.0$\pm$4.4 
& $14.18 \pm 0.02$ & 14.18$^{+0.11}_{-
0.09}$\tablenotemark{f} & $0.13^{+0.14}_{-0.12}$ & $-
0.66^{+0.14}_{-0.12}$ \\
   \                & 1526.71 & $-129\pm 1$ & 196.6$\pm$6.0 
& $14.15 \pm 0.04$ & 14.18$^{+0.11}_{-
0.09}$\tablenotemark{f} & $0.13^{+0.14}_{-0.12}$ & $-
0.66^{+0.14}_{-0.12}$ \\
Al~II\dotfill & 1670.79 & $-128\pm 1$ & 
222.8$\pm$10.0 & 12.98$\pm$0.04 & $13.02 \pm 0.04$ & $0.00 
\pm 0.11$ & $-0.80 \pm 0.11$ \\
Fe~II\dotfill & 1608.45 & $-131\pm 1$ & 79.3$\pm$5.4 
& 13.88$\pm$0.04 & 13.89$\pm 0.04$ & $-0.08 \pm 0.12$ & $-
0.60 \pm 0.12$ 
\enddata
\tablenotetext{a}{See Table~\ref{lineprop} for definitions 
of the quantities in this table. The equivalent widths and 
apparent column densities are integrated over the velocity 
range of the high-velocity component affiliated with the 
Outer Arm, from $v_{\rm LSR}$ = $-169$ to $-99$ km s$^{-
1}$.}
\tablenotetext{b}{Implied logarithmic abundance {\it if 
ionization corrections are neglected}, i.e., 
[X$^{i}$/H~I] = log $N$(X$^{i}$)/$N$(H~I) - 
log (X/H)$_{\odot}$.}
\tablenotetext{c}{Logarithmic abundance obtained by 
applying the ionization correction from the CLOUDY model 
shown in Figure~\ref{cloudy1821} and discussed in 
\S~\ref{secouter}. Error bars include column density 
uncertainties and solar reference abundance uncertainties 
but do not reflect uncertainties in the ionization 
correction. In this case, the ionization correction allows 
a large range of abundances (see \S~\ref{secouter}).}
\tablenotetext{d}{Saturated absorption line.}
\tablenotetext{e}{Due to considerable saturation, the 
formal error bars from Voigt-profile fitting may 
underestimate the true uncertainty.}
\tablenotetext{f}{Simultaneous fit to the Si~II 
$\lambda$1304.37 and $\lambda$1526.71 lines.}
\end{deluxetable}

\begin{figure}
\epsscale{1.0}
\plotone{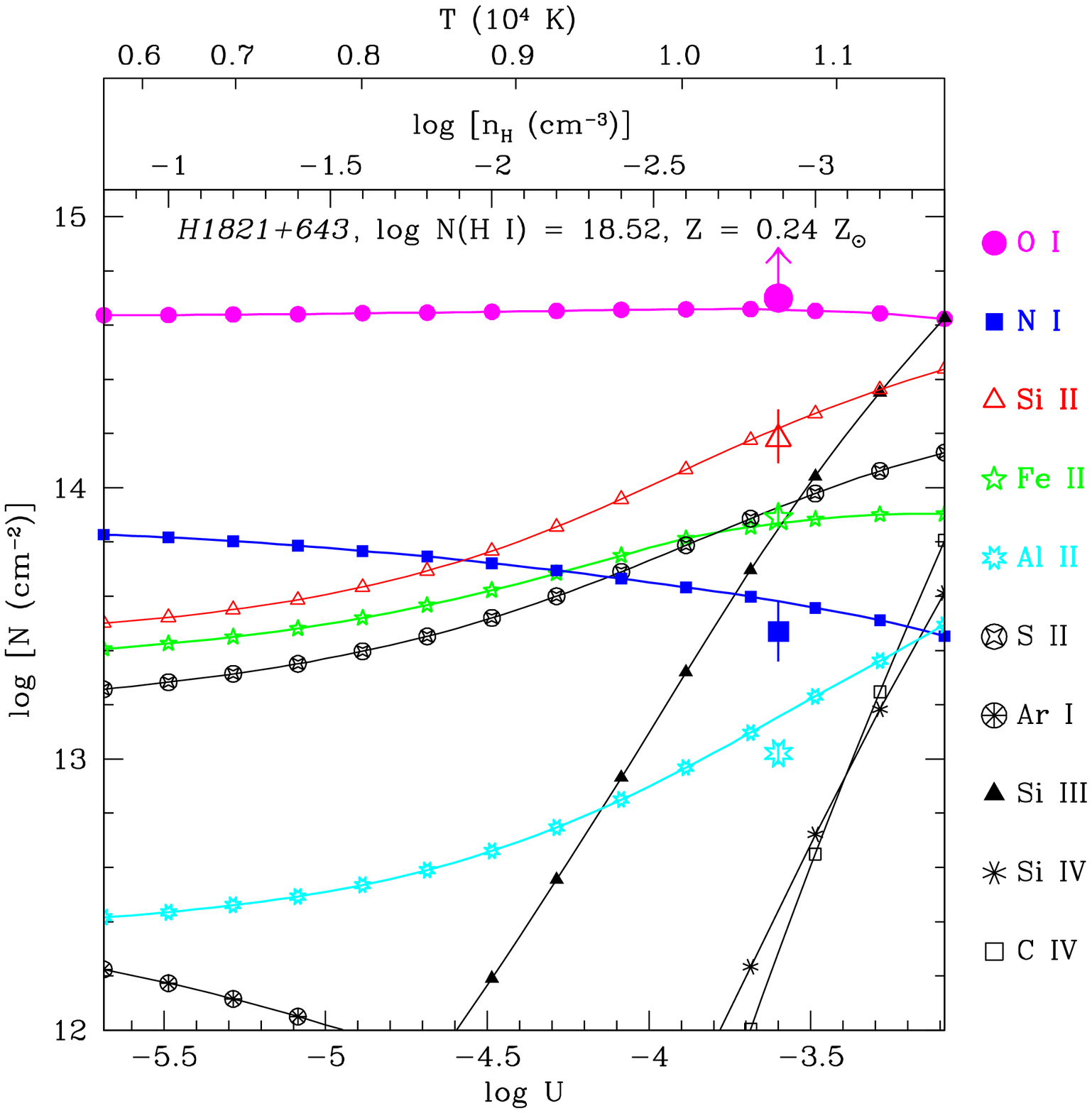}
\caption[]{Model of gas photoionized by the extragalactic
UV background, as in Figure~\ref{cloudymd}, but with
parameters appropriate for the Outer Arm component in the
spectrum of H1821+643. The large points show the observed
Outer Arm column densities toward H1821+643. At log $U = -
3.6$, the model is in reasonable agreement with the
observed column densities. However, $N$(\ion{O}{1}) may be
underestimated due to saturation; to fit the observations
with a higher \ion{O}{1} column density, $U$ must decrease
and the metallicity must increase.\label{cloudy1821}}
\end{figure}

\begin{deluxetable}{rcccll}
\tabletypesize{\footnotesize}
\tablewidth{0pc}
\tablecaption{High-Velocity Absorption Lines toward 
H1821+643: The -212 km s$^{-1}$ 
Component\tablenotemark{a}\label{hvr1821prop}}
\tablehead{\ \ Species \ \ & $\lambda _{0}$ & $<v>$ & 
$W_{\lambda}$ & \ \ log $N_{\rm a}$ & \ \ log $N_{\rm pf}$ 
\\
 \ & (\AA ) & (km s$^{-1}$) & (m\AA ) & \ & \ }
\startdata
O~I\dotfill & 1302.17 & \nodata & $3.8 \pm 4.7$ & $< 
13.38$\tablenotemark{b} & \nodata \\
C~II\dotfill & 1334.53 & \nodata & $7.6 \pm 3.4$ & 
$<12.84$\tablenotemark{b} & \nodata \\
Si~III\dotfill & 1206.50 & \nodata & $-3.9 \pm 8.6$ & 
$< 12.21$\tablenotemark{b} & \nodata \\
Si~IV\dotfill & 1393.76 & \nodata & $10.0 \pm 4.6$ & 
$< 12.32$\tablenotemark{b}  & \nodata \\
C~IV\dotfill & 1548.20 & $-212 \pm 1$ & $52.4 \pm 
5.5$ & $13.22 \pm 0.05$ & $13.24 \pm 0.03$\tablenotemark{c} 
\\
N~V\dotfill & 1238.82 & \nodata & $7.4 \pm 5.4$ & $< 
13.01$\tablenotemark{b}  & \nodata \\
O~VI\dotfill & 1031.93 & $-192$\tablenotemark{d} & 
\nodata & 13.72$\pm$0.14\tablenotemark{d} & \nodata
\enddata
\tablenotetext{a}{See Table~\ref{lineprop} for definitions 
of the quantities in this table. The equivalent widths and 
apparent column densities are integrated from $v_{\rm LSR}$ 
= $-234$ to $-184$ km s$^{-1}$.}
\tablenotetext{b}{4$\sigma$ upper limit derived from the 
equivalent width limit assuming the linear 
curve-of-growth applies.}
\tablenotetext{c}{Column density from a joint fit to the 
C~IV 1548.20 and 1550.78 \AA\ transitions.}
\tablenotetext{d}{O~VI line centroid and column 
density from Sembach et al. (2002).}
\end{deluxetable}
The larger equivalent widths toward H1821+643 at $v_{\rm 
LSR} \approx -130$ km s$^{-1}$ ostensibly indicate that the 
metallicity at this velocity is greater in the H1821+643 
component than in the 3C 351 HVC since the H~I 
columns are similar toward both QSOs. If we neglect 
ionization corrections, we indeed obtain high abundances 
from the column densities in Table~\ref{outerarmprop}, 
e.g., [\ion{O}{1}/\ion{H}{1}] $\gtrsim -0.6$, 
[\ion{Si}{2}/\ion{H}{1}] = 0.13, [\ion{Al}{2}/\ion{H}{1}] = 
0.0, and [\ion{Fe}{2}/\ion{H}{1}] = $-0.08$, but with a 
surprisingly low nitrogen abundance, 
[\ion{N}{1}/\ion{H}{1}] = $-0.97$. This would be an unusual 
abundance pattern; N underabundances are expected in 
low-metallicity gas, but as $Z \longrightarrow Z_{\odot}$, 
the contribution from intermediate-mass stars is expected to 
bring the N relative abundance more in line with the solar 
value. However, we have shown in \S~\ref{secabun} that 
ionization corrections must not be neglected when 
$N$(\ion{H}{1}) is at the level observed toward H1821+643. 
When ionization corrections are applied, the observed 
columns can be reconciled with a lower overall metallicity 
and smaller N underabundances.  Figure~\ref{cloudy1821} 
shows a CLOUDY model in satisfactory agreement with the 
observed columns with $Z = 0.24 Z_{\odot}$ and only small 
underabundance of N and Al. The model shown in 
Figure~\ref{cloudy1821} works if $N$(\ion{O}{1}) is close 
to the value from direct integration, i.e., $N$(\ion{O}{1}) 
$\approx$ 14.7. Higher \ion{O}{1} column densities would 
require higher metallicity, lower ionization parameters (to 
match, e.g., the \ion{O}{1}/\ion{Si}{2} ratio), and greater 
nitrogen underabundances, as can be seen from 
Figure~\ref{cloudy1821}. A slight depletion of Fe might 
also be required in higher metallicity models. Due to the 
large range of $U$ allowed by the current data, we cannot 
usefully bracket the size of the absorber at $v_{\rm LSR} = 
-130$ km s$^{-1}$.

At this juncture, it remains possible that the Outer Arm 
high-velocity gas and Complex C have similar metallicities, 
but the data also allow the Outer Arm to have a 
substantially higher metallicity than Complex C if N is 
underabundant. Consequently, it is difficult to draw firm 
conclusions about the relationship between the Outer Arm 
and Complex C.  Weaker \ion{O}{1} lines in the {\it FUSE} 
bandpass may help to more tightly constrain the metallicity 
of the Outer Arm and thereby clarify this relationship.

\begin{figure}
\plotone{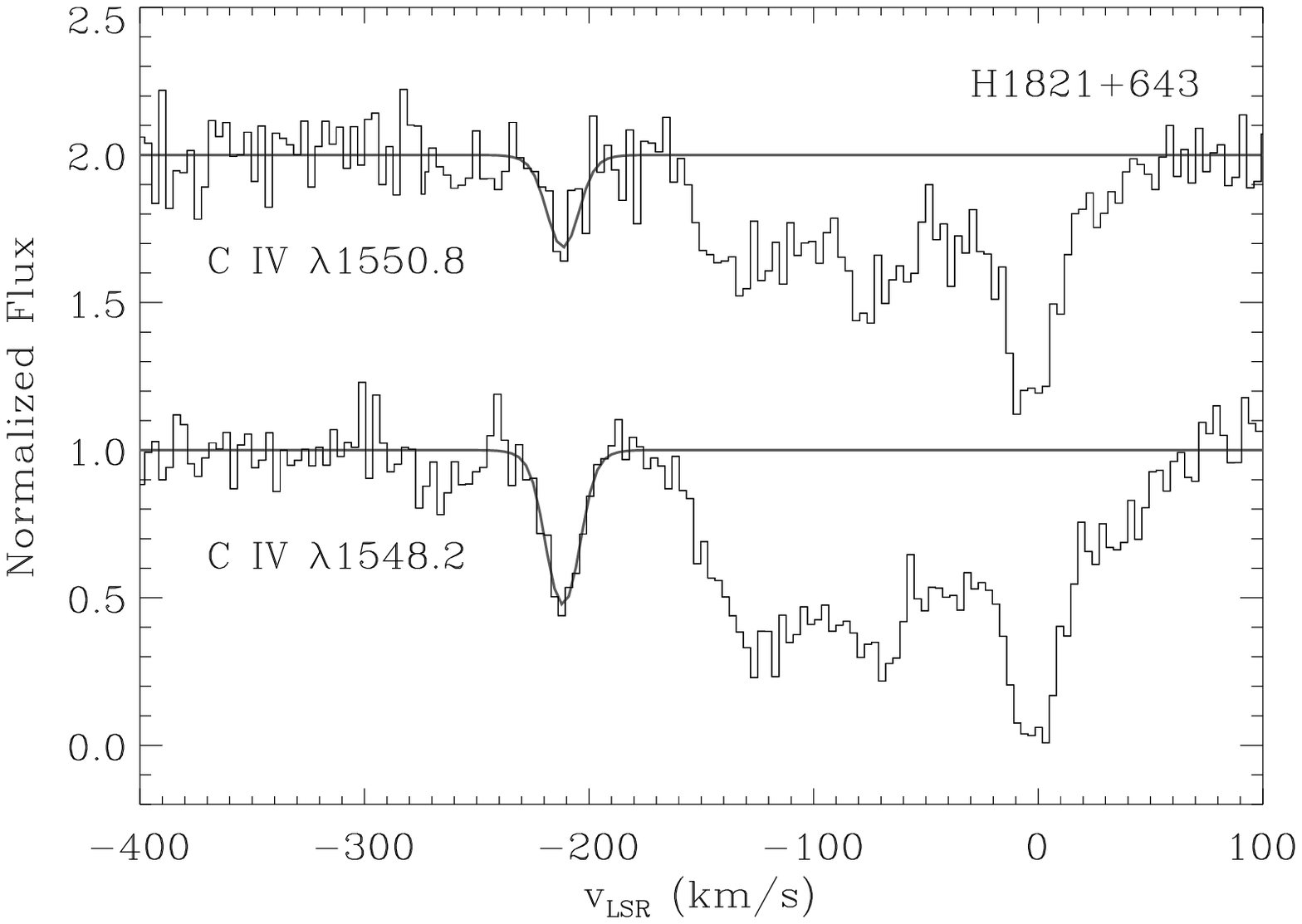}
\caption[]{Continuum-normalized \ion{C}{4} absorption
profiles observed toward H1821+643. The upper histogram
shows the $\lambda$1550.8 transition and the lower
histogram is the $\lambda$1548.2 line. The fit to the
component at $v_{\rm LSR} = -212$ km s$^{-1}$ is
overplotted with a solid line.\label{c4fit}}
\end{figure}

{\it The v = $-$200 km s$^{-1}$ component: the 
high-velocity ridge.}
Finally, it is interesting to compare the 3C 351 and 
H1821+643 sight lines at the velocity of the HVR component, 
$v_{\rm LSR} \approx -200$ km s$^{-1}$. While the HVR 
absorption is clearly detected in a variety of transitions 
toward 3C 351, most of the H1821+643 profiles show nothing 
at this velocity (the \ion{Si}{3} profiles in panel (e) of 
Figure~\ref{h1821compare} show the contrast between the 
sight lines at this $v$). However, Savage et al. (1995) 
tentatively identified \ion{C}{4} absorption near this 
velocity in a GHRS spectrum of H1821+643. As shown in 
Figure~\ref{c4fit}, we confirm this detection: both lines 
of the \ion{C}{4} $\lambda \lambda$1548.2, 1550.8 doublet 
are well-detected at $v = -212$ km s$^{-1}$. These are the 
only lines that we detect in the STIS spectrum at this 
velocity. The \ion{C}{4} equivalent width and column 
density measurements are listed in Table~\ref{hvr1821prop} 
along with upper limits on selected dominant ions and 
higher stages.  However, Oegerle et al. (2000) and Sembach 
et al. (2000) detected \ion{O}{6} at this velocity in {\it 
FUSE} spectra of H1821+643. The \ion{O}{6} 1031.9 \AA\ line 
is blended with low-velocity H$_{2}$, but Sembach et al. 
(2002) have removed the H$_{2}$ line (based on fits to 
other H$_{2}$ lines) and find log $N$(\ion{O}{6}) = $13.72 
\pm 0.14$.

These measurements have useful implications.  The fact that 
$N$(\ion{O}{6}) $\gg$ $N$(\ion{C}{4}) toward H1821+643 
would indicate that $T \gg 10^{5}$ K if the gas were 
collisionally ionized and in equilibrium (see, e.g., Figure 
7 in Tripp \& Savage 2000). However, this temperature is 
not compatible with the widths of the \ion{C}{4} lines. 
Figure~\ref{c4fit} shows our Voigt-profile fit to the HVR 
features in the H1821+643 spectrum; a single component with 
$b = 9\pm 2$ provides an excellent fit. This implies that 
$T < 10^{4.77}$ K. The \ion{C}{4} line width is only 
marginally consistent with the \ion{O}{6}/\ion{C}{4} ratio 
given the uncertainties in $N$(\ion{O}{6}) and 
$b$(\ion{C}{4}).  It seems more likely that either (1) much 
of the \ion{O}{6} absorption is not associated with the 
\ion{C}{4}-bearing gas (in which case the \ion{O}{6}-gas 
must be relatively hot to satisfy the lower limit on 
\ion{O}{6}/\ion{C}{4}), or (2) the gas is not in ionization 
equilibrium. The different centroids of the \ion{C}{4} and 
\ion{O}{6} lines (see Table~\ref{hvr1821prop}) support the 
idea that at least some of the \ion{O}{6} originates in 
different gas. It is unlikely that the \ion{C}{4} + 
\ion{O}{6} gas is photoionized because of the long 
pathlengths required (e.g., Sembach et al. 2002), but it is 
possible that the gas is collisionally ionized but out of 
equilibrium because it is cooling faster than it can 
recombine (Edgar \& Chevalier 1986). Heckman et al. (2002) 
have recently shown that a non-equilibrium, radiatively 
cooling gas model provides a good fit to the \ion{O}{6} 
column densities and $b-$values in HVCs (as well as other 
contexts). Their model also predicts $N$(\ion{O}{6}) $\gg$ 
$N$(\ion{C}{4}), as observed in the $-200$ km s$^{-1}$ 
feature toward H1821+643.

Savage et al. (1995) have noted that $v_{\rm LSR} \approx -
200$ km s$^{-1}$ is near the expected velocity for distant 
Milky Way gas in a corotating disk/halo in the direction of 
H1821+643; the velocity would then imply a large 
Galactocentric distance (see their Figure 2). However, it 
is quite possible that this high-velocity feature is 
associated with the high-velocity ridge observed toward 3C 
351 and other sight lines. In this case, the fact that the 
feature is only detected in \ion{C}{4} and \ion{O}{6} would 
indicate that the HVR has an extended ionized periphery. 
Evidence for highly-ionized layers on the surface of HVCs 
has been reported for other clouds (e.g., Sembach et al. 
1999). We have argued that the absorption lines toward 3C 
351 indicate that the HVR and Complex C proper are related. 
If the $v_{\rm LSR} = -212$ km s$^{-1}$ component toward 
H1821+643 is also part of the HVR, then this gas is likely 
much closer than implied by its velocity and the assumption 
that it is corotating (various arguments suggest that 
Complex C is $\sim$10 kpc from the Sun).

\section{Summary\label{secsum}}

We have investigated the physical structure, conditions, 
metallicity, and nature of the high-velocity cloud Complex 
C using high-resolution recordings of absorption lines in 
several directions through the cloud. Our study is mainly 
based on STIS echelle spectroscopy of 3C 351, which shows a 
wide variety of absorption lines from Complex C proper 
($v_{\rm LSR} = -128$ km s$^{-1}$) as well as an 
intermediate-velocity cloud (C/K, $v_{\rm LSR} = -82$ km 
s$^{-1}$) and a higher-velocity component that we refer to 
as the high-velocity ridge ($v_{\rm LSR} = -190$ km s$^{-
1}$). We also make use of other sight lines through Complex 
C from the literature as well as new STIS echelle 
observations of H1821+643, which is just outside the 21 cm 
boundary of Complex C but shows absorption lines at very 
similar velocities to those seen toward 3C 351. From our 
analysis, we reach the following conclusions:
\begin{enumerate}
\item The high-velocity ridge is closely related to Complex 
C proper. This is suggested by the similar morphologies of 
the HVR and Complex C; the HVR has a similar shape and is 
roughly centered on the larger, lower-velocity Complex C. 
This idea is supported by the absorption lines, which show 
remarkably similar column density ratios in the HVR and 
Complex C. However, very little high-ion absorption is 
detected at the velocity of Complex C proper, but 
\ion{Si}{3}, \ion{Si}{4}, and \ion{C}{4} absorption is 
strong and easily detected in the HVR. Moreover, the 
high-ion absorption lines in the HVR have centroids and 
line widths that are similar to those of the low ions in 
the HVR. We show that the high and low ionization stages in 
the HVR cannot arise in the same phase, and we conclude 
that the high ion absorption occurs in an interface between 
the low-ionization phase and a hotter ambient medium.
\item The relative abundances in all components with $N$(\ion{H}{1})
  $< 10^{19}$ cm$^{-2}$ indicate that ionization corrections are
  important. We consider collisional ionization equilibrium as well as
  CLOUDY photoionization models to derive ionization corrections, and
  we find in both cases that $Z = (0.2 \pm 0.1) Z_{\odot}$ in Complex
  C in the direction of 3C 351. We also find that nitrogen must be
  underabundant toward 3C 351. The low metallicity and nitrogen
  underabundance indicate that Complex C is not ejecta generated in a
  Galactic fountain. It seems more likely that Complex C is either
  tidally stripped material from a satellite galaxy (analogous to the
  Magellanic Stream), or that it is gas with a more distant
  extragalactic origin.  The absolute metallicity of the 3C 351 HVR
  component is less constrained because we only have an upper limit on
  the HVR \ion{H}{1} column. The lower limits on the HVR metallicity
  are consistent with the metallicity derived for Complex C proper.
\item We find similar oxygen abundances at Complex C 
velocities toward Mrk 279, Mrk 817, and PG1259+593. While 
there is some dispersion in the [O/H] measurements, within 
the current uncertainties the measurements are fully 
consistent with a constant metallicity throughout the HVC. 
These sight lines as well as the Mrk 876 sight line also 
provide strong evidence of N underabundances in Complex C. 
Comparison of the nitrogen abundances toward Mrk 876 and 
PG1259+593 provides the strongest evidence of spatial 
abundance variability in Complex C, but this requires 
confirmation with additional measurements. There are hints 
of $\alpha -$element overabundances in some directions, but 
the evidence is not statistically significant.
\item The derived iron abundances indicate that Complex C 
contains little or no dust.
\item Toward 3C 351, the Complex C absorber size, density, 
and temperature implied by the ionization models indicate 
that the gas is not gravitationally confined, while toward 
other sight lines such as PG1259+593, the implied path 
length through Complex C is much larger and gravitational 
confinement is viable. Pressure confinement by an external 
medium may play a role. However, we suggest that the 3C 351 
sight line passes through the leading edge of Complex C, 
which has been ablating and dissipating as it approaches 
the plane of the Milky Way. This idea is supported by 
\ion{O}{6} observations, which show the highest \ion{O}{6} 
column densities near the 3C 351 sight line. Furthermore, 
the \ion{O}{6}/\ion{H}{1} ratio increases dramatically with 
decreasing longitude and latitude within Complex C, which 
also suggests that the lower longitude and latitude regions 
are interacting more vigorously with the ambient medium.
\item To look for evidence of outflowing Milky Way gas that 
might interact with Complex C, we compare the STIS echelle 
observations of the sight lines to 3C 351 and H1821+643, 
which are both near the lower-latitude edge of Complex C at 
$l \approx 90^{\circ}$. We find that the 
intermediate-velocity gas observed toward both of these 
QSOs likely has a higher metallicity and may indeed be a 
fountain/chimney outflow from the Perseus spiral arm. 
Unfortunately, the results are less clear for the 
high-velocity gas toward H1821+643. The H1821+643 HVC could have 
a similar metallicity to that derived for Complex C, or it 
could have a much higher abundance depending on the 
\ion{O}{1} column and ionization correction, both of which 
are highly uncertain.
\end{enumerate}

\acknowledgements

As usual, we are grateful to Gary Ferland and collaborators 
for the development and maintenance of CLOUDY. The 
observations reported here were obtained by the STIS 
Investigation Definition Team, and this research was 
supported by NASA through funding for the STIS Team, 
including NASA contract NAS5-30110. TMT appreciates 
additional support for this work from NASA Long Term Space 
Astrophysics grant NAG5-11136 as well as NASA grant 
GO-08695.01-A from the Space Telescope Science Institute. 
BPW was supported by NASA grant NAG5-9024.

\end{document}